\documentclass[twocolumn]{IEEEtran} 
\usepackage[T1]{fontenc}
\usepackage{color}
\usepackage{mathrsfs}
\usepackage{amsmath}
\usepackage{amsthm} 
\usepackage{amssymb}
\usepackage{graphicx}
\usepackage{cite}
\usepackage{makecell}
\usepackage{algorithm}
\usepackage{algorithmic}
\usepackage{fancyhdr}
\usepackage{diagbox}
\usepackage{bm}
\usepackage{lettrine}
\usepackage{graphicx}

\usepackage[unicode=true,
bookmarks=true,bookmarksnumbered=true,bookmarksopen=true,bookmarksopenlevel=1,
breaklinks=false,pdfborder={0 0 0},pdfborderstyle={},backref=false,colorlinks=false]
{hyperref}
\hypersetup{
pdftitle={Your Title},
pdfauthor={Your Name},
pdfpagelayout=OneColumn, pdfnewwindow=true, pdfstartview=XYZ, plainpages=false}

\usepackage{multirow} 
\usepackage{xcolor}

\allowdisplaybreaks[4]

\usepackage[caption=false,font=footnotesize]{subfig}

\IEEEoverridecommandlockouts

\usepackage{geometry}
\usepackage{cleveref}
\begin{document}

%\pagenumbering{arabic}
\captionsetup[figure]{labelfont={},name={Fig.},labelsep=period}
\title{\textcolor{black}{UAV Swarm Deployment and Trajectory for 3D Area Coverage via Reinforcement Learning}}

\title{\fontsize{23pt}{25pt} \selectfont{Dynamic Trajectory Optimization and Power Control for Hierarchical UAV Swarms in 6G  Aerial Access Network }}
\author{\IEEEauthorblockN{  Ziye Jia,  \IEEEmembership{Member, IEEE,} Jia He, Lijun He, \IEEEmembership{Member, IEEE,} Min Sheng, \IEEEmembership{Fellow, IEEE,}  Junyu Liu, \IEEEmembership{Member, IEEE,}\\ Qihui Wu, \IEEEmembership{Fellow, IEEE,} and Zhu Han, \IEEEmembership{Fellow, IEEE }}
%\thanks{Ziye Jia is with the College of Electronic and Information Engineering, Nanjing University of Aeronautics and Astronautics, Nanjing 211106, China, and also with the State Key Laboratory of ISN, Xidian University, Xi'an 710071, China (e-mail: jiaziye@nuaa.edu.cn).}
\thanks{This work was supported  in part by National Natural Science Foundation of China under Grant 62301251 and 62201463, in part by the Natural Science Foundation on Frontier Leading Technology Basic Research Project of Jiangsu under Grant BK20222001, in part by the Aeronautical Science Foundation of China 2023Z071052007, and in part by the Young Elite Scientists Sponsorship Program by CAST 2023QNRC001. (Corresponding author: Min Sheng)}
\thanks{Ziye Jia, Jia He and Qihui Wu are with the College of Electronic and Information Engineering, Nanjing University of Aeronautics and Astronautics, Nanjing 211106, China (e-mail: jiaziye@nuaa.edu.cn; 071940128hejia@nuaa.edu.cn; wuqihui@nuaa.edu.cn).}
\thanks{Lijun He is with the School of Information and Control Engineering, China University of Mining and Technology, Xuzhou 221116, China (e-mail: lijunhe@cumt.edu.cn).}
\thanks{Min Sheng and Junyu Liu are with the State Key Laboratory of Integrated Services Networks (Xidian University), Xian, 710071,
 China (e-mail: msheng@mail.xidian.edu.cn; junyuliu@xidian.edu.cn).}
\thanks{Zhu Han is with the University of Houston, Houston, TX 77004 USA, and also with the Department of Computer Science and Engineering, Kyung Hee University, Seoul 446-701, South Korea (e-mail: hanzhu22@gmail.com).}
}

\maketitle

\begin{abstract}
%Unmanned aerial vehicles (UAVs) are widely applied in various applications due to the highly flexible, adaptable, and cost-efficient characteristics, which can act as aerial base stations (BSs) to extend the connectivity for ground users (GUs) in the sixth-generation (6G) era. However, the ability of a single UAV is limited and cannot support increasing complex tasks. Hence, in this paper, we investigate the cooperation of UAV swarms for ground data collection in the 6G aerial access networks (AAN). 
Unmanned aerial vehicles (UAVs) can serve as aerial base stations (BSs) to extend the ubiquitous connectivity for ground users (GUs) in the  sixth-generation (6G) era. 
However, it is challenging to cooperatively deploy multiple UAV swarms in large-scale remote areas.
Hence, in this paper, we propose a hierarchical UAV swarms structure for 6G aerial access networks, where the head UAVs serve as aerial BSs, and tail UAVs (T-UAVs) are responsible for relay.
In detail, we jointly optimize the dynamic deployment and trajectory of UAV swarms, which is formulated as a multi-objective optimization problem (MOP) to concurrently minimize the energy consumption of UAV swarms and GUs, as well as the delay of GUs. However, the proposed MOP is a mixed integer nonlinear programming and NP-hard to solve. 
Therefore, we develop a K-means and Voronoi diagram based area division method, and construct Fermat points to establish connections between GUs and T-UAVs.
Then, an improved non-dominated sorting whale optimization algorithm is proposed to seek Pareto optimal solutions for the transformed MOP. Finally, extensive simulations are conducted to verify the performance of proposed algorithms by comparing with baseline mechanisms, resulting in a 50\% complexity reduction.  
 %, which significantly upgrades the efficiency of data collection.
%In particular, we propose a hierarchical structure of UAV swarms, where the head UAV (H-UAV) serves as an aerial BS responsible for data processing, while the tail UAVs (T-UAVs) are deployed to collect and relay data from GUs in the 6G aerial access networks (AAN). 
\end{abstract}
\begin{IEEEkeywords}
6G aerial access network (AAN), UAV swarm trajectory planning,  multi-objective optimization problem (MOP), Voronoi diagram, Fermat point, improved non-dominated sorting whale optimization algorithm (INS-WOA).
\end{IEEEkeywords}

\section{Introduction}
\lettrine[lines=2]{W}{ith} the increasing development of the sixth generation (6G) wireless networks,  it is necessary to provide seamless and ubiquitous services for a massive number of ground users (GUs).
However, it lacks ground base stations (BSs) in most remote areas, which is challenging for realizing ubiquitous communication services \cite{b1,b7}.
Unmanned aerial vehicles (UAVs) are envisioned as a promising paradigm due to the mobility and flexibility \cite{b2,b4,markov},  which can serve as aerial BSs for both data collection and relay  in the 6G aerial access network (AAN) \cite{b5}. Moreover, since UAVs can fly close to GUs and establish low-altitude ground-to-air (G2A) communication links, they can be deployed to hover over the target areas, which can help save the energy cost and prolong the operational lifetime of GUs \cite{c1}. 
%Especially, a novel successive-hover-fly structure is presented for UAV to enable efficient trajectory design \cite{b16}. 
However, the limited capacity of a single UAV seriously constrains the ability to provide timely services for multiple GUs,  especially in large areas.

UAV swarms are proposed to provide large-area coverage services and improve the data collection efficiency in AAN, since the cooperation of multiple UAVs enables stronger capacities and flexibilities \cite{b3,b8}. 
Moreover, the transmission delay is a significant metric for time-sensitive GUs, such as in scenarios of emergency monitoring and safety protection \cite{c2,b6}. 
Consequently, the deployment and trajectory of UAV swarms are more intractable, in which the comprehensive consideration of UAV energy efficiency, GUs energy cost, and transmission delay of GUs is imperative \cite{b9}, \cite{Obstacles}. 
Besides, the balance among these metrics is significant. 
Therefore, how to satisfy the energy consumption and transmission delay requirements of remote GUs through flexible deployment of UAV swarms, dynamic adjustment of UAV trajectories and appropriate power control is a key issue.

To this end, we first propose a hierarchical framework for large-area data collection in AAN, with each swarm composed of a head UAV (H-UAV) and several tail UAVs (T-UAVs).
Then, to  jointly optimize the dynamic deployment and trajectory of UAV swarms, as well as the power control of GUs and T-UAVs,  we formulate the UAV swarms assisted data collection multi-objective optimization problem (USDC-MOP)  to simultaneously minimize the total energy consumption of  UAV swarms, the energy consumed by each GU, and the transmission delay of GUs. Since the proposed multi-objective optimization problem (MOP) is a mixed-integer nonlinear programming (MINLP) and NP-hard to solve, %which is NP-hard to solve \cite{b90},
we present the algorithm to facilitate the pre-deployment of UAV swarms by designing the K-means and Voronoi diagram based area division method, and Fermat points based connections establishment mechanism. 
Then, to efficiently tackle the transformed USDC-MOP with discrete and continuous variables, we propose an improved non-dominated sorting whale optimization algorithm (INS-WOA) to search for Pareto optimal solutions.

 The main contributions of this paper are summarized as follows.
\begin{itemize}
  \item [1)] 
  %We propose a hierarchical UAV swarm model composed of heterogeneous UAVs to accomplish AAN data collection in remote areas. Each UAV swarm consists of an H-UAV serving as the aerial BS and several T-UAVs responsible for data collection and relay. In this model, T-UAVs dynamically adjust trajectories to satisfy the transmission delay demands and ensure optimal coverage for remote GUs. 
  We propose a hierarchical UAV swarm model with H-UAVs as aerial BSs and T-UAVs for the data collection in remote areas. T-UAVs can dynamically adjust trajectories to meet transmission delay demands and ensure the optimal coverage for remote GUs.

  \item [2)] 
  %We focus on optimizing the deployment and trajectory of UAV swarms, as well as the power control of GUs and T-UAVs, which is formulated as a USDC-MOP to simultaneously minimize the total energy consumption of the UAV swarms, the energy consumed by GUs, and the transmission delay of GUs. The formulated USDC-MOP can also  clearly depict the balance among these significant metrics.
  We optimize the UAV swarm deployment, trajectories, and  power control for the GUs and T-UAVs. It is formulated as a USDC-MOP to minimize the energy consumption of UAV swarms and GUs, and transmission delay of GUs simultaneously. Additionally, it clearly illustrates the balance among these significant metrics.

  \item [3)] 
  %To efficiently solve the USDC-MOP, we firstly propose a pre-deployment algorithm for UAV swarms by utilizing the K-means and Voronoi diagram methods to divide the GUs into different regions for dynamic deployments.
  %For each region, we generate several Fermat points to establish connections between GUs and T-UAVs at different hovering positions.  
  %Furthermore, we propose an INS-WOA method to efficiently tackle the transformed USDC-MOP.
  To efficiently solve the USDC-MOP, we propose a UAV swarm pre-deployment algorithm that leverages K-means and Voronoi diagrams to partition GUs into regions for dynamic deployments. The fermat points are generated within each region to facilitate connections between GUs and T-UAVs across varying hovering positions. Additionally, an INS-WOA method is implemented to effectively tackle the transformed USDC-MOP.
  \item [4)]
  Extensive simulations are conducted to evaluate the performance of the proposed algorithms under various circumstances. 
  The results demonstrate their superiorities over benchmark MOP schemes in the effectiveness and low time complexity.
\end{itemize}

The rest of this paper is organized as follows. Related works are presented in  Section \ref{Section2}. In Section \ref{Section3}, we introduce the system model and formulate the USDC-MOP. Furthermore,  algorithms are designed in Section \ref{Section4}. Section \ref{Section5} conducts simulations and analyzes the results. Finally, conclusions are drawn in Section \ref{Section6}. 
\section{Related Works} \label{Section2}
\begin{figure*}
  \centering
  \begin{flushleft}
      \includegraphics[width=0.95\linewidth]{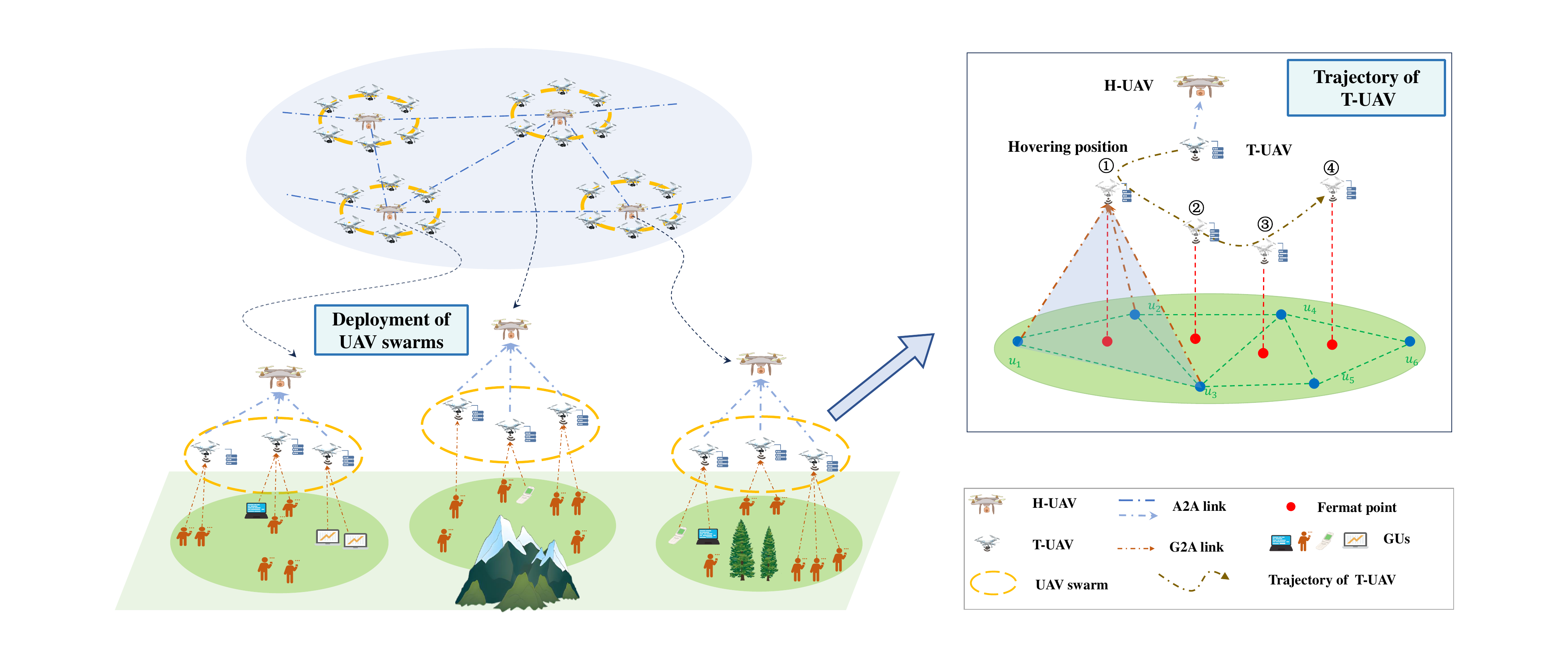}
  \caption{Hierarchical UAV swarms assisted 6G AAN data collection.}
  \label{fig1}
  \end{flushleft}
  \end{figure*}
  \vspace{1pt}

%Considering the recent research works, this section aims to provide a comprehensive introduction of AAN data collection assisted by single UAV, multiple UAVs, and UAV swarms, respectively. 
\subsection{Single UAV }
As for the single UAV assisted data collection, \cite{b16} presented a UAV-aided data collection framework to minimize the completion time from multiple sensors, where a novel successive-hover-fly structure was presented for the UAV. Meanwhile, to enable the efficient trajectory design, the convex approximation and an iterative algorithm were employed to jointly optimize the UAV trajectory and sensor assignment. The work of \cite{b17} considered an orthogonal frequency division multiple access (OFDMA) based UAV relay network, and the communication mode, sub-channel allocation, power allocation, as well as the UAV trajectory were jointly optimized to improve the quality of service of users. In \cite{b21}, a hybrid offline-online optimization scheme was developed for the UAV-enabled data harvesting scenario, which jointly designed the UAV trajectory and communication scheduling by leveraging both the statistical and instantaneous channel state information. It is observed that in the scenario of a single UAV, the limited energy capacity of the UAV restricts the performance of large-scale data collection.
\subsection{Multiple UAVs} 
For multiple UAV-assisted scenarios, \cite{b19} studied the problem of minimizing the total average age of information in a multi-UAV data collection system via a multi-agent deep reinforcement learning algorithm. Since the optimization for multiple UAVs trajectory was coupled with communication resource allocation, \cite{b20} considered the various mobility constraints and proposed an offline convex optimization method and an online convex-assisted reinforcement learning method to jointly optimize UAV trajectories, user association, and power control. 
The work of \cite{c8} studied the joint optimization of trajectory planning, communication design for multiple UAVs BSs, and access control of GUs, by presenting a multi-head attention mechanism to ensure efficient and fair communication.
\cite{Graph} proposed a novel graph attention multi-agent trust region reinforcement learning framework to solve the multi-UAV assisted communication problem, by introducing the graph recurrent network to process and analyze the complex topology of communication networks.
However, these works have not taken full advantages of the cooperation among UAVs for higher efficiency. Comparatively, UAV swarms are more applicable for complex missions in both military and civilian fields, due to the close collaboration and
flexible large-scale coverage.

\subsection{UAV Swarms}
Several recent works have begun to study the UAV swarm-enabled data collection. The authors in \cite{b22} discussed a UAV swarm assisted time-sensitive data collection system for remote sensors, highlighting the complexities in planning UAV swarm trajectories with constraints of  UAV capabilities and budget limitations. In work \cite{b27}, authors suggested the integration of formation flight and swarm deployment for multiple UAVs by designing a distributed control framework inspired by biological systems, enabling efficient formation and collision avoidance in dynamic environments. 
In \cite{c7}, the authors considered a scenario of UAV swarm deployment and trajectory to realize three-dimensional (3D) coverage with the effects of obstacles, and presented the Q-learning method to minimize the total trajectory loss of
the UAV swarm.
The work in \cite{b24} investigated a UAV swarm-assisted aerial-ground collaborative computing system to enhance the computation efficiency for ground smart mobile devices by optimizing the group formation, UAV trajectories, and resource allocation.  However, unlike homogeneous UAVs in \cite{b22,b27,c7,b24}, our hierarchical structure decouples base-station functions with H-UAVs and T-UAVs for reducing data delay, enabling the dynamic trajectory design in each swarm.

Although the above works have been studied towards UAV swarm-assisted data collection, most of them have not investigated the hierarchical swarm framework of heterogeneous UAVs for data collection. Furthermore, most studies focused on the deployment of the individual UAV swarm, without the consideration of the collaborative deployment among multiple UAV swarms. Besides, the energy efficiency and transmission delay of GUs are mostly ignored, which are crucial metrics for service qualities. 
Unlike the existing works focusing on the single swarm deployment, our hierarchical model enables the large-scale cooperative coverage, while the joint optimization of trajectory and power control addresses the trade-off between the energy efficiency and latency-critical services.
%Therefore, in this work, we focus on investigating a hierarchical UAV swarms assisted data collection model in AAN to tackle these issues. 

\section{System Model and Problem Formulation} \label{Section3}

As shown in Fig. \ref {fig1}, we consider a hierarchical UAV swarms assisted 6G AAN data collection scenario.  In particular, a set  $\mathcal{U}=\left\{1,2,\cdots, U\right\}$ of GUs are distributed in a remote large-scale area to sense the environment information, and a set $\mathcal{S}=\left\{1,2,\cdots,S\right\}$ of UAV swarms are deployed to provide data collection services. Specifically, any UAV swarm $s \in \mathcal{S}$ consists of one H-UAV $h_s \in \mathcal{H}=\left\{h_1,h_2,\cdots,h_S\right\}$ served as the aerial BS, and a set $\mathcal{V}_s=\left\{v_1^s,v_2^s,\cdots, V_{M_s}^s\right\}$ of T-UAVs responsible for collecting and relaying data for multiple GUs, where $v_{m}^{s}$ denotes the $m$-th T-UAV in UAV swarm $s$.  
%As shown in Fig. \ref {fig1}, we consider a hierarchical UAV swarms assisted 6G AAN data collection scenario, in which a set of GUs ${\mathcal{U}} = \{g_1,g_2,...,u\}$ with $u$ representing the $u$-th GU, are distributed in a remote large-scale area to sense the environment information, such as temperature, humidity, illumination, etc. Furthermore, $S$ hierarchical UAV swarms are deployed to provide data collection services. Specifically, each UAV swarm consists of one H-UAV served as the aerial BS and multiple T-UAVs responsible for collecting and relaying data for multiple GUs.
%For clarity, the set of $S$ H-UAVs is denoted as $\mathcal{H}=\{h_1,h_2,\cdots,h_S\}$ with $h_s$ representing the $s$-th H-UAV, and the set of all T-UAVs are defined as $\mathcal{V}=\{\mathcal{V}_1,\mathcal{V}_2,\cdots,\mathcal{V}_S\}$. Moreover, $\mathcal{V_S}$ denotes the set of T-UAVs assigned to the $s$-th UAV swarm, i.e., $\mathcal{V}_s=\{v_1^{s},v_2^{s},\cdots,v_{M_s}^{s}\}$,
Notably, $M_s$ is an integer variable denoting the number of T-UAVs in swarm $s$, which is optimized depending on the distribution of GUs. In other words, we can flexibly adjust the number of T-UAV $M_s$ in different UAV swarms to improve data collection efficiency, and the following constraints should be satisfied,
\begin{equation}
  \label{01}
  M_s \le M_{max}, \forall s,
\end{equation}
and
\vspace{-0.5em}
\begin{equation}
  \label{02}
  \sum\limits_{s=1}^S{M_s}= M,
\end{equation}
where $M_{max}$ and $M$ represent the maximum number of T-UAVs within a UAV swarm and total number of T-UAVs. The key notations used in this paper are summarized in Table \ref{notation}.
We employ a 3D Cartesian coordinate $\Theta$ to describe the data collection scenario,
i.e.,
\begin{equation}
  \Theta =\left\{   \left( x,y,z \right) \left| \begin{array}{c}
  X_{\min}\le x\le X_{\max}\\
  Y_{\min}\le y\le Y_{\max}\\
  Z_{\min}\le z\le Z_{\max}\\
\end{array} \right. \right\},
\end{equation}
where $\left(x,y,z\right)\in \Theta$ represents the 3D coordinate of a random position. Hence, 
the coordinate of GU $u$ is expressed as ${{q_u}} = \left({x_u},{y_u},{z_u}\right) $. Taking the real terrain into account, the heights of GUs are not uniform. The positions of H-UAV $h_s$ and T-UAV $v_{m}^{s}$ in UAV swarm $s$ are denoted by ${q_s}= \left({x_s},{y_s},{z_s}\right)$ and ${q_m^{s}}= \left({x_m^s},{y_m^s},{z_m^s}\right)$, respectively. The distance $d_{u,m}$ between GU $u$ and T-UAV $v_m^s$ is calculated as 

\begin{table}[tbp]
  \renewcommand\arraystretch{1.35}
  \caption{Key Notations }\label{notation}
  \begin{center}
    \fontsize{7.8}{10}\selectfont
    \begin{tabular*}{9cm}{@{\extracolsep{\fill}}|l|l|}
      \hline
      Notation & Description\\
      \hline
      $\mathcal{U}$, $\mathcal{S}$, $\mathcal{H}$ & Set of GUs, UAV swarms, H-UAVs\\
      $U$, $S$, $M$ & Number of GUs, UAV swarms, and T-UAVs \\
      ${q_u}$, ${q_s}$, ${q_m^s}$ & 3D Coordinates of GU $u$, H-UAV $h_s$, and T-UAV $v_{m}^s$\\
      $\mathcal{V}_s$ & Set of T-UAVs in UAV swarm $s$\\
      $M_s$ & Integer variable indicating the number of T-UAVs in swarm $s$\\
      $N$ & Number of hovering points of all T-UAVs\\
      $N_m^s$ & Number of hovering points of the T-UAV $v_{m}^s$\\
      $\phi_m^s$ & Trajectory path variable of T-UAV $v_{m}^s$\\
      $\Psi_m^s $ & Set of visiting sequences for T-UAV $v_{m}^s$\\
      $\gamma_{u,m,s}^n$ & Connection variable between GU $u$ and T-UAV $v_{m}^s$\\
      $d_{u,m}$ & Distance between GU $u$ and T-UAV $v_{m}^s$\\  
      $d_{m,s}$ & Distance between T-UAV $v_{m}^s$ and H-UAV $h_s$\\
      $h_{u,m}$ & Channel gain between GU $u$ and T-UAV $v_{m}^s$\\
      $h_{m,s}$ & Channel gain between T-UAV $v_{m}^s$  and H-UAV $h_s$\\
      $R_{u,m}^{tr}$ & Transmission rate from GU $u$ to T-UAV $v_{m}^s$\\
      $R_{m,s}^{tr}$ & Transmission rate from T-UAV $v_{m}^s$ and H-UAV $h_s$\\
      $p_u$ & Transmission power variable of GU $u$ \\ 
      $p_m^s$ & Transmission power variable of T-UAV $v_{m}^s$ \\ 
      $T_{u}$ & Transmission delay of GU $u$\\
      $T_{u,m}$ & Transmission delay of G2A \\
      $T_{m,s}$ & Transmission delay of A2A \\
      $T_{m}^{n}$ & Hovering duration of T-UAV $v_{m}^s$ at hovering position $q_m^{s}(n)$\\
      $T_{m}^{n,n+1}$ & Flight duration of T-UAV $v_{m}^s$ from two hovering positions \\
      $E_{m}$ &  Energy consumption of T-UAV $v_{m}^s$\\
      $E_{u}$ & Energy consumption of GU $u$\\
      $E_s$ & Energy consumption of H-UAV $h_s$\\
      \hline
    \end{tabular*}
  \end{center}
\end{table}
%We consider the three-dimensional (3D) Cartesian coordinate system. Each GU is positioned on the ground with 3D coordinates represented by ${q_u} = ({x_u},{y_u},{z_u}) \in \mathbf{R}^3, u \in {\mathcal{U}} = \{ 1,2,...,U\}$. Taking into account the practical complex terrain, the actual height of GUs is adjacent but distinct on the ground. On the other hand, the positions of H-UAV $s$ and T-UAV $v_{m}^s$ are  represented by ${q_s}= ({x_s},{y_s},{H}), s\in \mathcal{S}$ and ${q_m}= ({x_m},{y_m},{h_m}), m\in \mathcal{M}$, respectively. Unlike the leading H-UAV, which maintain a fixed flight altitude of $H$, relay T-UAVs are designed to fly freely during the data collection mission. Additionally, it is supposed that the specific locations of these GUs are known to UAVs in advance, such that it can perform the trajectory design. Moreover, the distance $d_{u,m}$ between the GU $v_{m}^s$ and T-UAV $v_{m}^s$ is stated as follows:
\hspace{-0.3em}
\begin{equation}\label{e1}
  \begin{split}
    {d_{u,m}} = \sqrt {{({x_m^s-x_u})^2} +{({y_m^s-y_u})^2} + {({z_m^s-z_u})^2}}.
  \end{split}
\end{equation}
Similarly, the distance $d_{m,s}$ between T-UAV $v_m^s$  and  H-UAV $h_s$ can be derived as
\begin{equation}\label{e111}
  \begin{split}
    {d_{m,s}} = \sqrt {{({x_m^s-x_s})^2} +{({y_m^s-y_s})^2} + {({z_m^s-z_s})^2} }.
  \end{split}
\end{equation}
\subsection{Deployment and Trajectory of UAV Swarms }
The deployment and trajectory of the UAV swarms are designed following the successive-hover-fly mode \cite{b16}. In other words, 
at the beginning, UAV swarms are deployed at locations $\bm{q_s}=\{q_s,\forall s\}$. Then, the T-UAVs depart from the swarm deployment positions and successively fly to their designated hovering locations to collect data from GUs, while the H-UAVs remain hovering at the deployment locations to receive data relayed from the T-UAVs, shown in Fig. \ref {fig1}. 
The trajectory path $\phi_m^s$ of T-UAV $v_m^{s}$ is described as the visiting sequence of $N_m^s$  hovering positions, i.e., $\phi_m^{s}=\{{q_m^{s}}(0),{q_m^{s}}(1),\cdots,{q_m^{s}}(N_m^s)\}$, where $q_m^s(n)$ is the $n$-th hovering positions of T-UAV $v_{m}^s$ ($n\in \{1,2,\cdots,N_m^s\}$), and the initial position is ${q_m^s}(0)={q_s}$. Further, the trajectory path $\phi_m^s$ is selected from the visiting sequence set $\Psi_m^{s}$, i.e., $\phi_m^s \in \Psi_m^{s}$.
Besides, the size of visiting sequence set $\Psi_m^{s}$ is related with the number of hovering positions $N_m^s$, which indicates that there are total $N_m^s!$ possible trajectory paths to be selected for each T-UAV $v_m^s$. The number of hovering locations for all T-UAVs is  $N=\sum\limits_{s=1}^{S}{\sum\limits_{m=1}^{M_s}{N_m^s}}$.

\subsection{Data Collection Model} 
As illustrated in Fig. \ref{fig1}, the channel models include the G2A model from GUs to T-UAVs, and the air-to-air (A2A) relay  model from T-UAVs to the H-UAV. To avoid interference during data collection, we assume that the OFDMA technique is adopted in both the G2A and A2A models, which implies that the T-UAV can simultaneously serve multiple GUs at the hovering points.
\subsubsection{G2A Channel Model}
%To manage the time-sensitive data collection tasks from GUs to relay T-UAVs and prevent interference among GUs during the collection process, we adopt an OFDMA scheme. We assume that the T-UAV must collect data from all GUs located within each hovering point. Thus, the binary variable $\gamma_{u,m}$ is introduced to represent the association status, i.e., if the GU $u$ is associated with the T-UAV $v_{m}^s$ , then $\gamma_{u,m}=1$, and otherwise, $\gamma_{u,m}=0$. To simplify the issue, it is assumed that each GU is only connected to one T-UAV at any given time, while each T-UAV has a capacity to serve up to $U_{max}$ GUs simultaneously. 

The G2A channel is considered as a probabilistic path loss model, which consists of both  line-of-sight (LoS) and non-LoS (NLoS) links with different probabilities. In particular, the probability of the LoS link  \cite {cc, eta} between GU $u$ and T-UAV $v_{m}^s$ is 
\begin{equation}\label{e4}
    {\cal P}_{u,m}^{{\rm{LoS}}} = \frac{1}{{1 + \alpha \exp \{  - \beta({\theta _{u,m}} - \alpha)\} }},\forall u, m,
\end{equation}
where $\alpha$ and $\beta$ are S-curve parameters depending on the environment, and ${{\theta _{u,m}}}$ is the elevation angle from GU $u$ to T-UAV $v_{m}^s$,
\begin{equation}\label{e5}
  \begin{split}
  {\theta _{u,m}}= \frac{{180}}{\pi }\arctan (\frac{z_{m}^{s}-z_{u}}{{\sqrt {({x_m^{s}} - {x_u})^2+({y_m^s} - {y_u})^2} }}), 
  \forall u, m.
  \end{split}
\end{equation}
Therefore, the G2A channel gain between GU $u$  and T-UAV $v_{m}^s$ is 
\begin{equation}\label{e6}
    {h_{u,m}} = {h_{u,m}^S}{h_{u,m}^F},\forall u,  m.
\end{equation}
Wherein, $h_{u,m}^S$ and $h_{u,m}^F$ respectively denote the channel power gain for the small-scale fading and free space fading, i.e.,
  \begin{equation}\label{e7}
    h_{u,m}^S = {\cal P}_{u,m}^{LoS}{\eta _{LoS}} + {\cal P}_{u,m}^{NLoS}{\eta _{NLoS}},\forall u, m,
  \end{equation}
  and
  \begin{equation}\label{e8}
    h_{u,m}^F = {\left( {\frac{c}{{4\pi f{d_{u,m}}}}} \right)^2},\forall u, m,
  \end{equation}
where ${\cal P}_{u,m}^{NLoS} = {1-{\cal P}_{u,m}^{LoS}}$. ${\eta _{LoS}}$ and ${\eta _{NLoS}}$ indicate the mean additional losses for LoS and NLoS links, respectively. $f$ is the carrier frequency.
According to the Shannon formula, the G2A channel capacity from GU $u$ to T-UAV $v_{m}^s$ is 
\begin{equation}\label{e9}
  R_{u,m}^{tr} = {B_{u,m}}{\log _2}\left( 1+  {\frac{{{h_{u,m}}{p_u}}}{{{B_{u,m}}\sigma^2}}} \right),
  \forall u, m ,
\end{equation}
where $B_{u,m}$ denotes the  bandwidth between GU $u$ and T-UAV $v_{m}^s$, $\sigma^2$ represents the average noise power spectrum density, and ${p_u}$ is the transmission power of GU $u$. Note that the communication uplink is established only if the transmission rate  satisfies the following constraint:
\begin{equation}\label{e10}
  R_{u,m}^{tr} \geq R_{min}^{tr}, \forall u, m ,
\end{equation}
where $R_{min}^{tr}$ represents the threshold of transmission rate.

\subsubsection{A2A Channel Model}
The data collected by T-UAVs should be promptly relayed to the hovering H-UAVs for further processing.
%Due to the the limitations of energy capacity of T-UAVs, they have to forward the collected data to the leading H-UAV which acts as aerial BS for further processing. We assume that T-UAV forward the data as soon as completing gathering all the data at service points. The functions of relay T-UAVs for forwarding collected data are designed in an OFDMA manner that ensures non-interference between them, thereby facilitating smooth and efficient communication among T-UAVs. 
%Thus, the binary variable $b_{m,s}\in\{0,1\}$ represents the connection relationship between T-UAV $v_{m}^s$ and H-UAV $s$, which is also the indicator for the connectivity of swarm. Taking into account the energy constraints of the H-UAV, each swarm can accommodate up to $M_{max}$ data collection T-UAVs.
%Thus, the set of all T-UAVs served by the leading H-UAV $s$ is denoted as $\mathcal{M}^s=\{1,2,...,M_s\}$, where ${M_s}$ is the number of T-UAVs in swarm $s,s \in \mathcal{S}$ , and ${M_s} \le M_{max}$.  
The  NLoS fading are ignored due to the wider view of the A2A model. Hence, the A2A transmission rate from T-UAV $v_{m}^s$ to H-UAV $h_s$ is 
\begin{equation}
  R_{m,s}^{tr} = {B_{m,s}}{\log _2}\left( { 1+ \frac{{{\eta_{LoS}}{p_m^s}}}{{{B_{m,s}}\sigma^2}}} \right),
  \forall m, s,
\end{equation}
where $p_m^s$ represents the transmission power of T-UAV $v_{m}^s$, and $B_{m,s}$ is the bandwidth between T-UAV $v_{m}^s$ and H-UAV $h_s$. Similarly, the transmission rate is expected to satisfy the following constraint
\begin{equation}\label{e15}
  R_{m,s}^{tr} \geq R_{s,min}^{tr}, \forall m ,s,
\end{equation}
where $R_{s,min}^{tr}$ represents the threshold of transmission rate between the T-UAV and H-UAV.

%Further, the corresponding energy consumption used by relay T-UAV $v_{m}^s$ to forward the data is expressed as 
%\begin{equation}\label{e16}
 % E_{m,s}^{tr} = {p_m}{t_{m,s}}={p_m}\frac{\sum\limits_{u=1}^{U_k} {{Q_{u,m}}}}{R_{m,s}^{tr}},
  %\forall m,s,
%\end{equation}
%where $t_{m,s}$ and $U_k$ represent the required time for data forwarding and the number of GUs in cluster $C_k$, respectively.  

\begin{figure}
  \centering
    \includegraphics[width=0.43\textwidth]{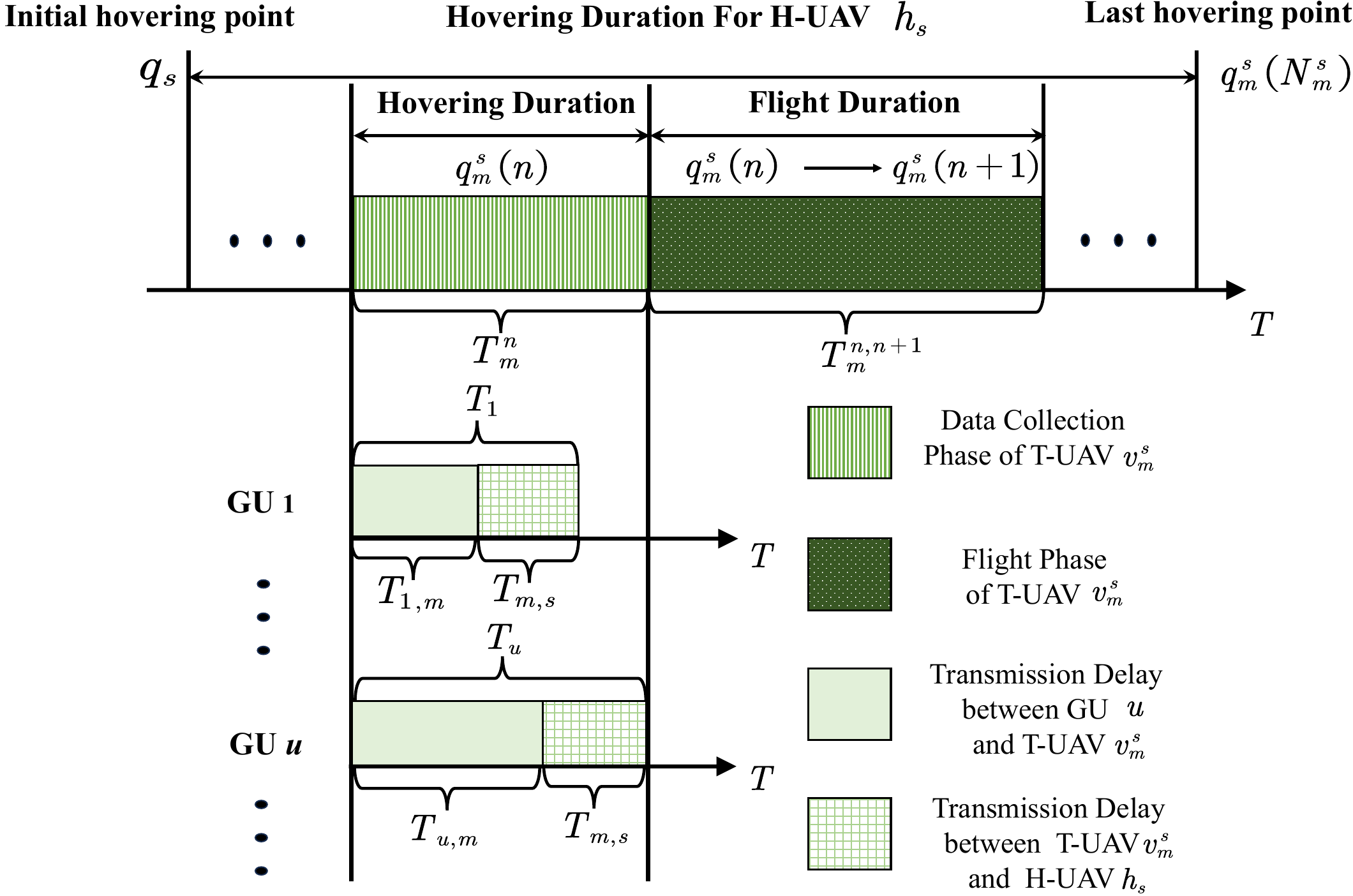}
  \caption{Time sequence of T-UAV data collection.}
  \label{time}
\end{figure}

\subsection{Delay Model}
The delay model  of T-UAV $v_m^s$ and related H-UAV $h_s$ is illustrated in Fig. \ref{time}.
Let binary variable $\gamma_{u,m,s}^n \in \{0,1\}$ denote the connection relationship between GU $u$ and T-UAV $v_{m}^{s}$ at the hovering location $q_m^s(n)$, i.e.,
\begin{equation}
  \gamma_{u,m,s}^{n}=\begin{cases}
    1, &\text{if GU }u\text{ is connected to T-UAV }v_m^s \text{ at } q_m^s(n), \\
    0, &\text{otherwise}.
  \end{cases}
\end{equation}

The implementation process of T-UAV includes two phases: the data collection phase and  flight phase. As for the data collection phase, the transmission delay $T_{u}$ of GU $u$ at hovering position $q_m^s(n)$ is calculated as 
\begin{equation}
  T_u=\sum\limits_{s=1}^S{\sum\limits_{m=1}^{M_s}{\sum\limits_{n=1}^{N_m^s}{\gamma_{u,m,s}^n(T_{u,m}+T_{m,s})}}}, \forall u,
\end{equation}
where $T_{u,m}$ and $T_{m,s}$ are the transmission delays of G2A and A2A communication links, respectively:
\begin{equation}
  T_{u,m}=\frac{Q_u}{R_{u,m}^{tr}}, \forall u, m,
  \label{tum}
\end{equation}
and
\begin{equation}
  T_{m,s}=\frac{Q_u}{R_{m,s}^{tr}}, \forall u, m, s,
\end{equation}
where $Q_{u}$ is the data size of GU $u$. 
$Q_{u}$ significantly influences the transmission delay, and as data volume of GUs increases, the transmission power rises proportionally.
Nevertheless, due to the disparity in transmission delays of different GUs, the hovering duration $T_{m}^{n}$ of T-UAV $v_{m}^s$  on the $n$-th hovering position should be no less than the longest transmission delay among these GUs, as illustrated in Fig. \ref{time}, which is calculated as
\begin{equation}
  T_{m}^{n}=\max_{u\in \left\{ 1,2,\cdots ,U \right\}}\left\{ \gamma _{u,m,s}^{n}\left( T_{u,m}+T_{m,s} \right)  \right\} ,\forall s, m, n.
\end{equation}

After finishing data collection and relay at position $q_m^s(n)$, T-UAV $v_{m}^s$ flies to the next hovering position $q_m^s(n+1)$. The time cost of T-UAV $v_{m}^s$ from ${q_m^s}(n)$ to ${q_m^s}(n+1)$ is 
\begin{equation}\label{e17}
T^{n,n+1}_{m}=\frac{||{q_m^s}(n)-{q_m^s}(n+1)||}{||{\vartheta||}},\forall m, n,
\end{equation}
where $||{q_m^s}(n)-{q_m^s}(n+1)||$ represents the Euclidean distance between ${q_m^s}(n)$ and ${q_m^s}(n+1)$. ${\vartheta}$ is the uniform velocity of each T-UAV, denoted by ${\vartheta}=\left( \vartheta_{x},\vartheta_{y},\vartheta_{z} \right)$. Accordingly, it is assumed that all UAVs fly at the same velocity during data collection. 
Moreover, H-UAV $h_s$ keeps hovering at the position ${q_s}$ until the last T-UAV $v_m^s$ complete data collection. Different flight velocities affect the flight time of T-UAVs, thereby influencing the flight energy consumption of the entire UAV swarm. In this paper, we focus on optimizing the multi-objective problem under the same flight velocity. 
Thus, the time cost of H-UAV $h_s$, consisting of all the  following T-UAVs from initial hovering point $q_{s}$ to the last hovering point $q_{m}^{s}$,  is calculated as
\begin{equation}
T_s=\max_{m\in \left\{ 1,2,\cdots,M_s \right\}} \left\{ \sum_{n=1}^{N_m^s}{T_{m}^{n}}+\sum_{n=0}^{N_m^s-1}{T_{m}^{n,n+1 }} \right\}, \forall s.
\end{equation}

In summary, variable $\gamma_{u,m,s}^n = 1$ can force T-UAV $v_{m}^{s}$ to hover at $q_{m}^{s}(n)$ until the GU's data is delayed. Therefore, to minimize the transmission delay $T_u$, we can adopt the trajectory design, by optimizing hovering positions $q_{m}^{s}(n)$ to reduce hovering durations $T_{u,m}$ in  (\ref{tum}), and sequencing waypoints $\phi_m^s$ to reduce flight durations in  (\ref{e17}).

\subsection{ Energy Consumption Model  }
Evaluating the energy consumption of UAV swarms and GUs is crucial in assessing the performance of AAN data collection. Therefore, we present the energy consumption models of UAV swarms and GUs, respectively.   

\subsubsection{Energy Consumption of UAV Swarms} 
The energy consumption of UAVs is mainly composed of the propulsion and communication \cite{b28}. As for the T-UAV,  the energy consumption primarily consists of three components: the flight energy, hovering energy, and communication energy. Specifically, the energy consumption $E_{m}$ of the T-UAV $v_{m}^s$ during the data collection task is calculated as
\begin{equation}
E_{m}=E_{m}^{tr}+E_{m}^{hov}+E_{m}^{fly},
\forall m,
\end{equation}
where $E_{m}^{tr}$ denotes the relay energy consumption of T-UAV $v_{m}^s$,
\begin{equation}
  E_{m}^{tr} = {p_m^s}\sum\limits_{u=1}^{U}{\sum\limits_{n=1}^{N_m^s}{\gamma_{u,m}^nT_{m,s}}},
  \forall m, s.
\end{equation}
Wherein, $p_{m}^s$ is the transmission power of T-UAV $v_{m}^s$. Further, $E_{m}^{hov}$ and $E_{m}^{fly}$ represent the hovering and flight energy consumption of T-UAV $v_{m}^s$, respectively. Moreover, we leverage the  successive-hover-fly model for the UAV, i.e., T-UAV is unable to communicate with either GUs or the H-UAV during the flight.
Considering the 3D scenario, the energy consumption models for horizontal and vertical movements of the UAV are different. Hence, the power consumption for the UAV flying in a straight-and-level manner \cite{b29} with the horizontal velocity
${\vartheta}_{x,y}=\left(\vartheta_x,\vartheta_y\right)$ is calculated as
\begin{align}
  \label{Pmov}
  & P_{fly}(||{\vartheta}_{x,y}||) =  P_0\biggl(1 + \frac{3||{\vartheta}_{x,y}^2||}{\mathbb{U}_{tips}^2}\biggr) \notag \\
  & + P_1\left(\sqrt {1 + \frac{||{\vartheta}_{x,y}^4||}{4v_0^4}} - \frac{||{\vartheta}_{x,y}^2||}{2v_0^2}\right)^{\frac{1}{2}}  
  + \frac{1}{2}d_0 \rho_0 s_0 A_0 ||{\vartheta}_{x,y}^3||,
\end{align}
where $P_0$ and $P_1$ denote the blade profile power and induced power in hovering status, respectively. $\mathbb{U}_{tips}$ is the tip speed of the rotor blade, and $v_0$ represents the mean rotor induced velocity. $d_0$ is the fuselage drag ratio, $\rho_0$ is the air density, $s_0$ is the rotor solidity, and $A_0$ is the rotor disc area. Further, when ${\vartheta}$ is 0, the hovering power consumption can be calculated as
\begin{equation}
  P_{hov}=P_0+P_1.
\end{equation}
The vertical flight  power consumption of the T-UAV is 
\begin{equation}
  P_{ver}(||{\vartheta}_{z}||)=W\mathbf{g}||{\vartheta}_{z}||,
\end{equation}
where $W$  is the mass of the UAV, $\mathbf{g}$ is the gravitational acceleration, and ${\vartheta}_{z}$ is the UAV vertical velocity.
Thus, $E_{m}^{hov}$ and $E_{m}^{fly}$ can be respectively calculated as 
\begin{equation}
  E_{m}^{hov} = {P_{hov}}\sum\limits_{n=1}^{N_m^s}{T_{m}^{n}},
  \forall m,
\end{equation}
and
\vspace{-0.5em}
\begin{align}
  E_{m}^{fly} = (P_{fly}(||{\vartheta}_{x,y}||)+P_{ver}(||{\vartheta}_{z}||)) 
  \sum\limits_{n=0}^{N_m^s-1}{T_m^{n,n+1}}, \forall m.
\end{align}

Furthermore,  H-UAV $h_s$ remains hovering at the deployment point $q_s$ during the data collection. Hence, the energy consumption of H-UAV $h_s$ is calculated as
\begin{equation}
E_s=P_{hov} T_s,
\forall s. 
\end{equation}

\subsubsection{Energy Consumption of GUs}
Based on the G2A channel model, the energy consumption $E_{u}$ of GU $u$ is mainly for data transmission to T-UAV $v_{m}^s$, which is calculated as 
\begin{equation}\label{e11}
  E_{u} = {p_u}\sum\limits_{s=1}^S{\sum\limits_{m=1}^{M_s}{\sum\limits_{n=1}^{N_m^s}{\gamma_{u,m,s}^nT_{u,m}}}},
  \forall u,
\end{equation}
where $p_{u}$ is the transmission power of GU $u$.
\subsection{Problem Formulation}

We formulate the USDC-MOP  to cooperatively  minimize the total energy consumption of UAVs swarms, the energy consumed by GUs, and transmission delay of GUs, by jointly optimizing the deployment of UAV swarms $\bm{q_s}=\{q_s,\forall s\}$,  the number of T-UAVs in UAV swarms $\bm{M_s} =\{M_s,\forall s\}$, the trajectory path of T-UAVs $\bm{\Phi}=\{\phi_{m}^s, \forall s, m \}$,  as well as the transmission power $\bm{P}=\{p_u,p_m^s, \forall u, s, m\}$ of GUs and T-UAVs. Further, the connection variable set is  $\bm{\gamma} =\{\gamma_{u,m,s}^{n}, \forall u,  m, s, n\}$. 
Thus, all variables of the problem are summarized  as $\bm{\mathbb{A}}=\left\{\bm{q_s},\bm{M_s},\bm{\Phi},\bm{P},\bm{\gamma}\right\}$ \cite{Marler}, and  three optimization objectives are illustrated as 
\begin{equation}
  \label{mop}
  \begin{cases}
     f_1\left( \mathbb{A} \right) =&\sum\limits_{s=1}^S{\left( E_s+\sum\limits_{m=1}^{M_s}{E_m} \right) },\\
     f_2\left( \mathbb{A} \right)= &\frac{1}{U}\sum\limits_{u=1}^U{E_u},\\
     f_3\left( \mathbb{A} \right)= &\frac{1}{U}\sum\limits_{u=1}^U{T_u}.
  \end{cases}
\end{equation}
Wherein, $f_1(\mathbb{A})$ is the total energy consumption of the UAV swarms (TEU), which is critical for evaluating the effectiveness of the UAV swarms assisted data collection. $f_2(\mathbb{A})$ represents the average energy consumption of GUs (AEG).  $f_3(\mathbb{A})$ is the average transmission delay of GUs (ADG). The total objective of the USDC-MOP is to simultaneously minimize TEU, AEG, and ADG, which is formulated as
\begin{subequations}
  \begin{align}
    \textbf{P0:}  \quad  \min_{\mathbb{A}}& \enspace F(\mathbb{A})=\left[f_1(\mathbb{A}),f_2(\mathbb{A}),f_3(\mathbb{A})\right]\notag \\
     {\rm{ s}}{\rm{.t.}}& \enspace (\ref{01}),(\ref{02}), (\ref{e10}),(\ref{e15}),  \notag\\
     &\quad \sum_{s=1}^S{\sum\limits_{m=1}^{M_s}{\sum\limits_{n=1}^{N_m^s}{\gamma_{u,m,s}^n}}}=1,\forall u,\label{006}\\
     &\quad \sum\limits_{u=1}^{U}{\gamma_{u,m,s}^n} \leq U_{max},\forall s, m, n, \label{008}\\
     &\quad T_u\le T_u^{\max},\forall u,\label{007}\\
     &\quad \gamma_{u,m,s}^{n} \in \{0,1\},\forall u, m, n,\label{005}\\
     &\quad {q_s} \in \Theta, \forall s,\label{00s}\\  
     &\quad \phi _m^s \in \Psi_m^s,\forall s, m, n,\label{002}\\ 
     &\quad p_u\in \left[ P_u^{\min},P_u^{\max} \right],\forall u,\label{003}\\ 
     &\quad p_m^s\in \left[ P_m^{\min},P_m^{\max} \right],\forall s, m.\label{004} 
   \end{align}
\end{subequations}
%\begin{subequations}
 % \begin{align}
  %  \textbf{P0:}  \quad  \min_{\mathbb{A}}& 
   % \begin{cases}
    %    f_1(\mathbb{A})= \sum\limits_{s=1}^S{\left( E_s+\sum\limits_{m=1}^{M_s}{E_m} \right) }\notag \\
     %  f_2(\mathbb{A})= \frac{1}{U}\sum\limits_{u=1}^U{E_u}\notag \\
      %  f_3(\mathbb{A})=\frac{1}{U}\sum\limits_{u=1}^U{T_u} \notag \\
 % \end{cases}\notag \\
  %   {\rm{ s}}{\rm{.t.}} &\quad (\ref{01}),(\ref{02}), (\ref{e10}),(\ref{e15}),  \notag\\
   %  &\quad \sum_{s=1}^S{\sum\limits_{m=1}^{M_s}{\sum\limits_{n=1}^{N_m^s}{\gamma_{u,m,s}^n}}}=1,\forall u,\label{006}\\
    % &\quad \sum\limits_{u=1}^{U}{\gamma_{u,m,s}^n} \leq U_{max},\forall s, m, n, \label{008}\\
     %&\quad T_u\le T_u^{\max},\forall u,\label{007}\\
     %&\quad \gamma_{u,m,s}^{n} \in \{0,1\},\forall u, m, n,\label{005}\\
     %&\quad {q_s} \in \Theta, \forall s,\label{00s}\\  
     %&\quad \phi _m^s \in \Psi_m^s,\forall s, m, n,\label{002}\\ 
     %&\quad p_u\in \left[ P_u^{\min},P_u^{\max} \right],\forall u,\label{003}\\ 
     %&\quad p_m^s\in \left[ P_m^{\min},P_m^{\max} \right],\forall s, m.\label{004} 
   %\end{align}
%\end{subequations}
Wherein, constraint (\ref{006}) specifies that each GU can only establish the connection with one T-UAV at one hovering location $q_m^s(n)$, and (\ref{008}) indicates that each T-UAV can simultaneously serve $U_{max}$ GUs. Constraint (\ref{007}) denotes the maximum tolerated delay  $T_{u}^{max}$ of GU $u$.
The maximum tolerated delay $T_{u}^{max}$ varies across GUs depending on their time-sensitive nature, which imposes additional constraints on the coverage capability and data collection efficiency of UAV swarms.
Constraints (\ref{00s})  restricts the deployment spatial range of H-UAVs and T-UAVs, respectively. Constraint (\ref{002}) limits the trajectory path according to the set of all visiting sequences $\Psi_m^s$ of T-UAV $v_{m}^s$. Constraints (\ref{003}) and (\ref{004}) indicate the range of transmit power for GU $u$ and T-UAV $v_{m}^s$, respectively.

Due to the binary connections variable between GUs and T-UAVs, the integer variable for the allocation of T-UAVs, and the non-linear form of the objective functions, problem $\textbf{P0}$ is an MINLP problem \cite{109,2009}, which is generally NP-hard to solve \cite{b90}. 
%Additionally, we consider a special case of problem, where several T-UAVs associated with the same  H-UAV fly above mission areas according to trajectory routes $\mathbf{\Phi}$. The determined routes can make the 3D path of the T-UAVs shortest, so that minimizing the energy consumption of UAVs with a fixed capacity. Therefore, the problem is essentially the same as the capacitated vehicle routing problem (CVRP), known as an NP-hard \cite{b30,b31}. 
Furthermore, these three optimization objective functions in $\textbf{P0}$ must be optimized simultaneously, and a balanced trade-off exists among these objectives, making $\textbf{P0}$ more intractable to figure out. Therefore, multi-objective optimization algorithms are regarded as the preferred algorithms for solving MOPs. However, $\textbf{P0}$ is an MOP with  various discrete and continuous variables due to the the existence of integer variables $\bm{M_s}$, $\bm{\gamma}$ and $\bm{\Phi}$, and thus, we simplify the original $\textbf{P0}$ with the Voronoi diagram and Fermat points based pre-deployment method 
and propose the INS-WOA to handle the transformed USDC-MOP.

\vspace{-0.5em}
\section{ Algorithm Design}\label{Section4}

\begin{figure}
  \centering
   \includegraphics[width=0.48\textwidth]{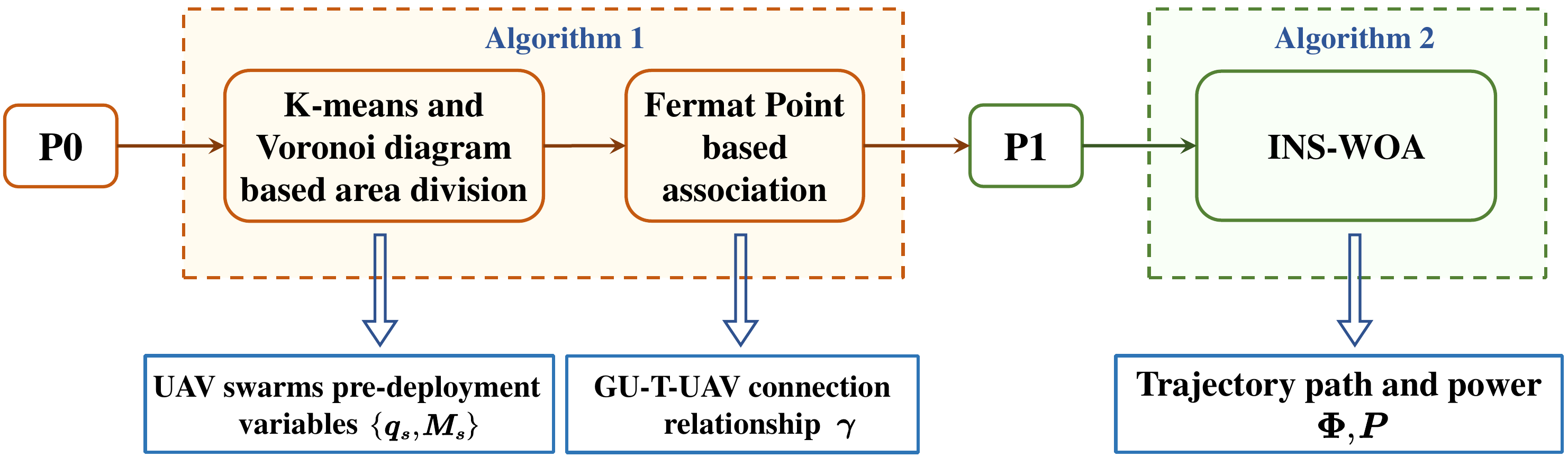}
  \caption{Overview of the designed algorithms.}
  \label{INS-WOA}
\end{figure}

To tackle $\textbf{P0}$ efficiently, we present two sequential algorithms, as depicted in Fig. \ref{INS-WOA}. Wherein, Algorithm \ref{A1} is designed for the pre-deployment of UAV swarms, in which the K-means and Voronoi diagram based area division method and Fermat point based connection method are proposed to obtain the key variables of the pre-deployment positions $\bm{q_s}$ of swarms, number of T-UAVs $\bm{M_s}$ assigned to $s$-th UAV swarm, and connections $\bm{\gamma}$.
Thus, the original problem $\textbf{P0}$ can be transformed into $\textbf{P1}$ with variables of $\{\bm{\Phi},\bm{P}\}$, which is a small-scale MOP.
%without variables $\{\bm{q_s},\bm{M_s},\bm{\gamma}\}$ after performing the proposed Algorithm \ref{A1}. % Besides, to simplify the search range, the value range for continuous variable $q_s$ in constraint (\ref{00s}) is discretized into $\Omega$, where $\Omega = \{\omega_{1},\omega_{2},\cdots,\omega_{\varsigma }\}$ denotes the set of possible deployment locations.%Thus, the dimensionality of $\textbf{P0}$ is decreased via pre-deployment operations. It is observed that $\textbf{P1}$ is still a typical MOP with variables of $\{\bm{\Phi},\bm{P}\}$, in which multiple objectives must be optimized simultaneously. 
Then, we propose the INS-WOA to deal with $\textbf{P1}$ via introducing a greedy  selection mechanism, and obtain the Pareto solutions set.

\vspace{-0.8em}
\subsection{Pre-deployment Algorithm for UAV swarms}\label{Section3a}

Generally, the effective deployment of UAV swarms can significantly enhance the efficiency of data collection. However, it is intricate to select the deployment locations $\bm{q_s}$ from the large task areas $\Theta_s$, which seriously enlarges the search space of the algorithm. GUs can be effectively clustered by the K-means algorithm. However, the cluster centers are not the ideal deployment locations for swarms. 
Therefore, we propose to divide the area based on the GU distributions leveraging the Voronoi diagram, with UAV swarms deployed at the intersections of diagrams, i.e., $\Omega = \{\omega_{1},\omega_{2},\cdots,\omega_{\varsigma }\}$, where $\omega_{\varsigma}$ denotes the intersection of the Voronoi diagram. Then, we can assign $M_s$ T-UAVs to the $s$-th UAV swarm, and the inner T-UAVs are dispatched to their respective subregions to complete data collection. As supplementary, the Voronoi diagram is a widely used partitioning mechanism in mathematics, computational geometry, and spatial analysis \cite{c10}. It can divide a plane or space into several regions, where each point within a region is closer to a specific point (referred to as the generator or seed point). Such partitioning is based on the nearest neighbor relationships. Therefore, we construct a two-dimensional (2D) Voronoi diagram based on K-means method for GUs partitions. 

Besides, constraint (\ref{006}) illustrates that T-UAVs are required to complete the data collection tasks for all GUs.
Since we have already determined the deployment locations of UAV swarms and the GUs that each T-UAV needs to serve, then we should determine at which specific hovering positions the T-UAVs establish connections with GUs. However, we are unable to determine the specific locations of the hovering points of T-UAVs. Fermat points \cite{c11} can ensure a shortest sum distance from the hovering point to vertices in a Delaunay triangle constructed
by GUs.
Therefore, we deploy the method of generating Fermat points to represent the pre-deployment hovering positions and establish connection relationships $\bm{\gamma}$ with the GUs.

In detail, the pre-deployment of UAV swarms is presented in Algorithm \ref{A1}. Firstly, we cluster the GUs $\mathcal{U}$ into $M$ clusters by the K-means method \cite{c20} to obtain the seed points $\mathcal{C}$ (step \ref{ll1}). Then, we construct the 2D Voronoi diagram $\mathcal{W}$ with seed points $\mathcal{C}$, as well as the cluster center points (step \ref{ll2}), as illustrated in Fig. \ref{3D}.  Subsequently, we calculate the 2D coordinates of the Voronoi vertices and obtain the pre-deployment locations $\bm{q_s}$ with the fixed deployment altitude of UAV swarms from $\Omega$ (step \ref{ll2}). Further, we obtain the number of T-UAVs $\bm{M_s}$ assigned to different UAV swarms according to constraints (\ref{01}) and (\ref{02}).
Based on Voronoi diagram  $\mathcal{W}$, we obtain Voronoi subregions $\mathcal{L}=\left\{L_1,L_2,\cdots,L_M\right\}$.
Furthermore, we generate $N_m^s$ Fermat points within each subregion using the geometric median method. For instance, we can generate the Fermat points for three GUs according to $\underset{\mathcal{F}_{m}^{s}\left( n \right)}{\min}\sum_{u=1}^3{\lVert \mathcal{F}_{m}^{s}\left( n \right) -q_u \rVert} $.
$H_n$ GUs are accordingly assigned to their nearest Fermat points, shown in Fig. \ref{3D}.  Then, the connection relationships $\bm{\gamma}$ between GUs and T-UAVs are determined. Therefore, the variables $\{\bm{q_s},\bm{M_s},\bm{\gamma}\}$ of $\textbf{P0}$ are obtained.

\begin{algorithm}[t]
  \caption{Pre-deployment of UAV Swarms } 
  \label{A1}
  \renewcommand{\algorithmicrequire}{\textbf{Input:}}
  \renewcommand{\algorithmicensure}{\textbf{Output:}}
  \begin{algorithmic}[1]
    \REQUIRE Locations of GUs $\mathcal{U}$, number of UAV swarms $S$, and number of total T-UAVs $M$.
    \STATE \textbf{Initialize} $M_s=0$, $\gamma_{u,m,s}^n=0$, and maximum number of iterations $I_{max}$.
    \STATE Cluster $\mathcal{U}$ into $M$ clusters using K-means method to obtain cluster centers $\mathcal{C}=\{C_1,C_2,\cdots,C_M\}$.\label{ll1}
    \STATE Construct Voronoi diagram $\mathcal{W} $ using centers $\mathcal{C}$ to obtain the set of Voronoi subregions  $\mathcal{L}=\left\{L_1,L_2,\cdots,L_M\right\}$ and  intersections of diagrams $\Omega = \{\omega_{1},\omega_{2},\cdots,\omega_{\varsigma  }\}$.\label{ll2}
    \STATE Select $\bm{q_s}$ from the set of intersections $\Omega$ and obtain the $\bm{M_s}$ due to constraints (\ref{01}) and (\ref{02}).
    \FOR { $s =1,2,\cdots, S$ }
      \FOR{$m =1,2,\cdots, M_s$}
        \FOR{$n =1,2,\cdots, N_m^s$}
        \STATE Generate Fermat points $\mathcal{F}_m^s(n)$ according to  and connect the GU $u$ to T-UAV $v_{m}^s$ at the hovering location $q_{m}^s(n)$, i.e., $\gamma_{u,m,s}^n=1$ according to constraints (\ref{006}) and (\ref{008}).
        \ENDFOR
      \ENDFOR
    \ENDFOR
    \ENSURE Voronoi diagram $\mathcal{W}$, pre-deployment positions  of UAV swarms $\bm{q_s}$, number of T-UAVs $\bm{M_s}$ assigned to each UAV swarm, and connection relationships $\bm{\gamma}$.
  \end{algorithmic}
\end{algorithm}

\subsection{INS-WOA Design}
After the implementation of Algorithm \ref{A1}, $\textbf{P0}$ is turned into $\textbf{P1}$ as
\begin{subequations}
  \begin{align}
    \textbf{P1:}  \quad  \min_{\mathbb{B}}& \enspace F'(\mathbb{B})=\left[f'_1(\mathbb{B}),f'_2(\mathbb{B}),f'_3(\mathbb{B})\right]\notag \\
  {\rm{ s}}{\rm{.t.}} &\quad  (\ref{007}),(\ref{002})-(\ref{004}),(\ref{e10}), (\ref{e15}).
   \end{align}
\end{subequations}
%\begin{subequations}
 % \begin{align}
  %  \textbf{P1:}  \quad \min_{\mathbb{B}}& 
   % \begin{cases}
   % f'_1\left( \bm{\mathbb{B}} \right) =\sum\limits_{s=1}^S{ E_s } +\sum\limits_{m=1}^{M}{E_m}  \notag \\
   % f'_2\left( \bm{\mathbb{B}} \right)= \frac{1}{U}\sum\limits_{u=1}^U{\left(p_uT_{u,m}\right)} \notag \\
   % f'_3\left( \bm{\mathbb{B}} \right)= \frac{1}{U}\sum\limits_{u=1}^U{\left(T_{u,m}+T_{m,s}\right)} \notag \\
   % \end{cases}\notag \\
   %  {\rm{ s}}{\rm{.t.}} &\quad (\ref{007}),(\ref{002})-(\ref{004}),(\ref{e10}), (\ref{e15}).
   % \end{align}
%\end{subequations}
%In detail, the original multi-objective functions in (\ref{mop}) are  transformed into:
%From Algorithm \ref{A1}, we obtain the pre-deployment variables $\{\bm{q_s},\bm{M_s,\bm{\gamma}}\}$. Thus, the original multi-objective functions \ref{mop} are transformed as:
Wherein, 
\begin{equation}
  \label{mop2}
  \begin{cases}
    f'_1\left( \bm{\mathbb{B}} \right) =\sum\limits_{s=1}^S{ E_s } +\sum\limits_{m=1}^{M}{E_m},  \\
    f'_2\left( \bm{\mathbb{B}} \right)= \frac{1}{U}\sum\limits_{u=1}^U{p_uT_{u,\bar{m}}}, \\
    f'_3\left( \bm{\mathbb{B}} \right)= \frac{1}{U}\sum\limits_{u=1}^U{\left(T_{u,\bar{m}}+T_{\bar{m},\bar{s}}\right)}. 
  \end{cases}
\end{equation}
are obtained according to the original multi-objective functions in (\ref{mop}), and, $\mathbb{B}=\{\bm{\Phi},\bm{P}\}$ denotes the set of unresolved variables.  Since the connections variable $\bm{\gamma}$ is obtained by Algorithm \ref{A1}, $\bar{m}$ and $\bar{s}$ in (\ref{mop2}) are determined by $u$. 
However, $\textbf{P1}$ is still an MINLP problem, and the traditional mathematical methods are unable to balance multiple conflicting objectives.  Hence, to tackle $\textbf{P1}$ efficiently, we design the INS-WOA to obtain variables $\bm{\Phi}$ and $\bm{P}$.

The traditional WOA is a bio-inspired optimization technique \cite{c12}, drawing inspiration from the unique hunting strategies of humpback whales. WOA updates the positions of solutions by emulating three primary hunting strategies of these whales, including encircling prey, bubble-net attacking method, and search for prey.  The NS-WOA incorporates a non-dominated sorting (NDS) mechanism, crowding distance calculation and sorting mechanism to solve complex MOP. However, $\textbf{P1}$ is intractable to solve with the discrete variable $\bm{\Phi}$ by using the NS-WOA. Thus,  we propose an INS-WOA, where a greedy selection mechanism  is introduced to select the visiting sequence on the basis of the NS-WOA.
The detailed INS-WOA to solve $\textbf{P1}$ is designed as follows.

\subsubsection{NDS Mechanism}
NDS is a pivotal mechanism for multi-objective optimization. It classifies the solutions into different non-dominated fronts, known as Pareto fronts (PF) \cite{c30}, where no solution within the same front dominates any other. The mechanism aids WOA in identifying efficient solutions for MOP, ensuring that the obtained solution set balances the performance across all objective functions in \textbf{P1}.

\begin{figure}[t]
  \centering
    \includegraphics[width=0.4\textwidth]{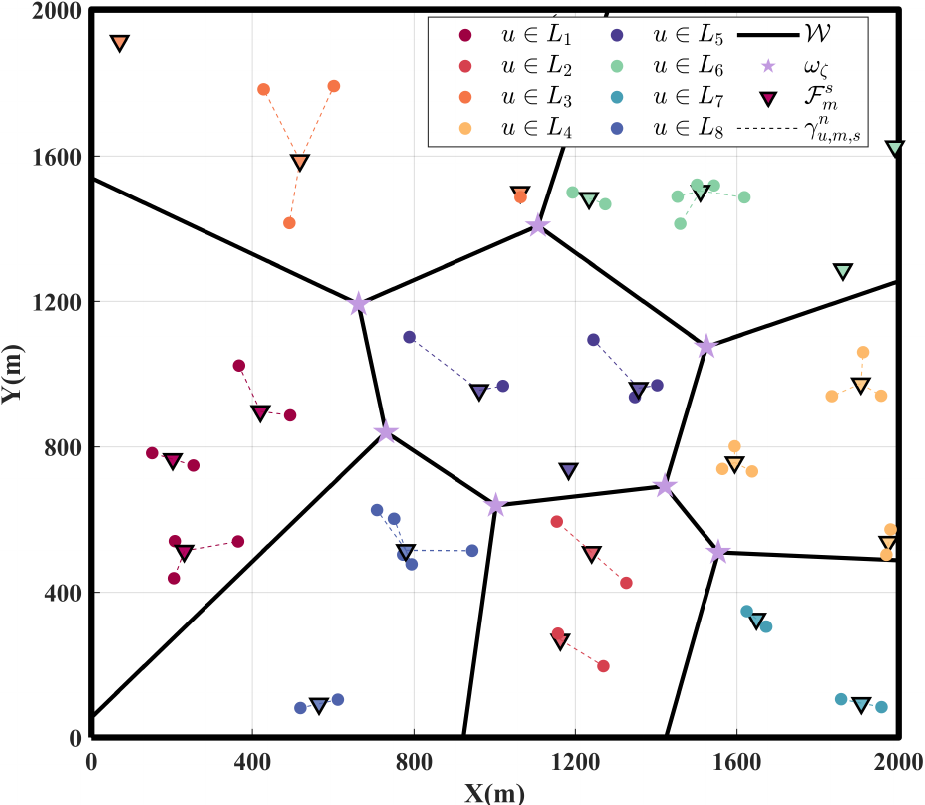}
  \caption{Area partitioning based on the Voronoi diagram,  including the connection relationship between the Fermat points and GUs.}
  \label{3D}
\end{figure}

\subsubsection{Crowding Distance Calculation and Sorting}
The purpose is to ensure that the solutions are uniformly distributed along the PF, rather than clustered together in specific regions.
The crowding distance for each solution is calculated as the sum of normalized distances between neighboring solutions in each objective dimension. A larger crowding distance indicates a more isolated solution, which is preferred to maintain diversity in the solution set.

\subsubsection{Encircling Prey}
The NS-WOA simulates the behavior of whales encircling their prey by adjusting the positions of individual whales to progressively obtain an optimal solution  \cite{b32}. This encircling behavior promotes the exploration and exploitation processes within the NS-WOA, facilitating the development of high-quality solutions, i.e.,
 \begin{equation}\label{z3}
  \vec{D}=\left| \vec{C}\odot \overrightarrow{X^*}\left( i \right) -\vec{X}\left( i \right) \right|,
 \end{equation}
 and
 \begin{equation}\label{z4}
  \vec{X}\left( i+1 \right) =\overrightarrow{X^*}\left( i \right) -\vec{A}\odot \vec{D},
 \end{equation}
where $i$ is the current iteration, $\vec{D}$ denotes the position of the target prey, $\overrightarrow{X^*}\left( i \right)$ is the position of the best search agent, $\odot$ denotes the the element-wise multiplication, and $\left| \odot\right| $ is the absolute value.
Further, $\vec{A}$ and $\vec{C}$ are coefficient vectors calculated as
 \begin{equation}\label{z5}
  \vec{A}=2\vec{a}\odot \vec{r}-\vec{a},
 \end{equation}
 and
 \begin{equation}\label{z6}
  \vec{C}=2\odot \vec{r},
 \end{equation}
where parameter $\vec{a}$ is linearly decreased from 2 to 0 during the iterations, governing both the exploration and exploitation phases. Parameter $\vec{r}$ is a random vector uniformly distributed within the range $[0, 1]$.  Let $I'_{max}$ denote the maximum number of iterations. Accordingly, parameter $\vec{a}$ is updated according to $\vec{a}=2(1-i/I'_{max})$.
The primary objective of  (\ref{z5}) and (\ref{z6}) is to strike a balance between exploration and exploitation for USDC-MOP solutions. Parameter $\vec{r}$ is randomly generated in both equations, and introduces randomness into the position updating mechanism of the agent population. Such randomness helps diversify the search process, preventing premature convergence and ensuring that NS-WOA can effectively explore the solution space while also exploiting unknown solutions.

\subsubsection{Bubble-Net Attacking}
The shrinking encircling and spiral updating position mechanisms are employed concurrently to simulate the bubble-net hunting strategy of humpback whales. To replicate the helix-shaped movement characteristic of humpback whales, the spiral equation defining  the relationship between the prey's location and the whale can be expressed as 
\begin{equation}\label{z7}
  \overrightarrow{D'} = \left| \overrightarrow{X^*}(i) - \overrightarrow{X}(i) \right|,
\end{equation}
in which
\begin{equation}\label{z8}
  \vec{X}(i+1) = \overrightarrow{D'} \odot e^{bl} \odot \cos(2\pi l) + \vec{X}^*(i),
\end{equation}
where $b$ is a constant used to define the logarithmic spiral shape, and $l$ is a random number within $[-1,1]$.

Since humpback whales simultaneously swim around their preys in a shrinking circle and follow a spiral-shaped path, the shrinking encircling method and the spiral approach are employed concurrently in the model. To accurately represent this behavior, it is assumed that each mechanism is executed with a probability of $50 \% $ as 
\begin{equation}\label{z9}
  \vec{X}(i+1)=\begin{cases}
      \overrightarrow{X^*}(i) - \vec{A} \odot \vec{D}, &\text{if } \tau   <0.5,\\
      \overrightarrow{D'} \odot e^{bl} \cos(2\pi l) + \overrightarrow{X^*}(i),&\text{if } \tau \ge 0.5,
  \end{cases}
\end{equation}
where $\tau$ is a parameter within $[0,1]$.
  
\subsubsection{Search for Prey}
The similar approach used in the shrinking encircling mechanism can be applied to the prey search process. Besides, the coefficient vector $\vec{A}$ with  $\vec{A}>1$ is employed, and the position $\overrightarrow{X^*}\left( i \right)$ of the best search agent is replaced by the position $\overrightarrow{X_{rand}}$ of a randomly selected whale from the current population. This adjustment forces the humpback whales to move away from a randomly chosen whale, thereby enabling the NS-WOA algorithm to expand the search space and conduct a global search. The mathematical model for the prey search can be expressed as

\begin{equation}\label{z10}
  \vec{D}=\left| \vec{C}\odot \overrightarrow{X_{rand}} -\vec{X}\left( i \right) \right|,
 \end{equation}
 and
 \begin{equation}\label{z11}
  \vec{X}\left( i+1 \right) =\overrightarrow{X_{rand}}-\vec{A}\odot \vec{D}.
 \end{equation}

 \subsubsection{Greedy Selection Mechanism}
 In the process of solving optimization problems, it usually needs to go through a series of steps, and at each step, multiple choices are faced. However, the greedy selection  mechanism makes the nearly optimal choice at present step when solving problems.  As for $\textbf{P1}$, a greedy mechanism is utilized to select the next hovering positions for T-UAVs from the solutions obtained by NS-WOA, with the selection based on minimizing a weighted sum of  objective changes, i.e.,
 \begin{equation}\label{greedy}
 \Delta F'(\mathbb{B}) = \frac{\Delta f'_1(\mathbb{B})}{f'_1(\mathbb{B})^*} + \frac{\Delta f'_2(\mathbb{B})}{f'_2(\mathbb{B})^*} +  \frac{\Delta f'_3(\mathbb{B})}{f'_1(\mathbb{B})^*}.
\end{equation}
Wherein, $\Delta f_{j}(\mathbb{B})$ is the predicted change in objective $j$, and ${f'_j(\mathbb{B})^*}$ is the current best value.

\begin{algorithm}[!t]
  \caption{INS-WOA for $\textbf{P1}$}
  \label{A2}
  \renewcommand{\algorithmicrequire}{\textbf{Input:}}
  \renewcommand{\algorithmicensure}{\textbf{Output:}}
  \begin{algorithmic}[1]
  \REQUIRE Locations of GUs $\mathcal{U}$, Voronoi diagram $\mathcal{W}$, pre-deployment positions $\bm{q_s}$, number of assigned T-UAVs $\bm{M_s}$, and connection relationships $\bm{\gamma}$.
  %\STATE \textbf{Initialize} $X_{min}, X_{max}, Y_{min}, Y_{max}, H_{min}, H_{max}, P, P^\prime$,\\
  %$U, S, M, M_s, H, K, \Omega, \varphi, q_u, q_s^{ini}, q_m^{ini}, f_1, f_2, f_3$.
  \STATE \textbf{Initialize} the population $\mathbf{X}$ of whales agents, number of whales agent $J_{max}$, maximum number of iterations $I'_{max}$, and the control parameters $\vec{A}$, $\vec{C}$, $\tau$, and $l$. \label{l1}
    \FOR{ each hovering position $n=0$ to $N_m^s$}
      \STATE Evaluate the fitness of each whale in the population.\label{l6}
      \STATE Apply NDS to classify the population into fronts.\label{l7}
      \STATE Compute crowding distances for solutions within each front.\label{l8}
      \STATE Select the best solution $\mathbf{X}^*$  based on the rank and crowding distance.\label{l9}
      \FOR{each iteration $i = 1$ to $I'_{max}$}
        \FOR{each whale $j = 1$ to $J_{max}$}
          \STATE Update parameters $\vec{A}$, $\vec{C}$, $\tau$, and $l$.\label{l10}
          \IF{$p < 0.5$}
            \IF{$|\vec{A}| < 1$}\label{l11}
              \STATE Update $\vec{D}$ and  $\overrightarrow{X}\left( i \right)$ due to (\ref{z3}) and (\ref{z4}), respectively.
            \ELSE
              \STATE Select a random whale $\overrightarrow{X_{rand}}$ and update $\vec{D}$ according to (\ref{z9}).
            \ENDIF
          \ELSE
            \STATE Update  $\overrightarrow{D'} $ via (\ref{z10}) and  $\vec{X}(i+1)$ via  (\ref{z8}).
          \ENDIF \label{l12}
        \ENDFOR
        \STATE Repeat steps \ref{l6}-\ref{l9}. \label{l13}
    \ENDFOR  
    \STATE  Select the next hovering position $q_m^s(n+1)$ for T-UAV $v_m^s$ according to the greedy selection mechanism that minimizes (\ref{greedy}).\label{l3}
  \ENDFOR 
  \ENSURE The PFs solutions  $\bm{\Phi},\bm{P}$ and MOP objectives $f'_1(\mathbb{B})$, $f'_2(\mathbb{B})$, $f'_3(\mathbb{B})$ of $\textbf{P1}$.

  \end{algorithmic}
\end{algorithm}

In detail, the INS-WOA for $\textbf{P1}$ is presented in Algorithm \ref{A2}, the input includes locations of GUs $\mathcal{U}$, Voronoi diagram $\mathcal{W}$, the pre-deployment positions of UAV swarms $\bm{q_s}$, the assigned number of T-UAVs $\bm{M_s}$, and connection relationships $\bm{\gamma}$. Then, the related parameters are initialized, encompassing key aspects of the population of  whale agents  $\mathbf{X}$, along with setting the maximum number of iterations $I'_{max}$ and the control parameters of the NS-WOA  $\vec{A}$, $\vec{C}$, $\tau$,  and $l$. 
Subsequently, we utilize the NS-WOA method to determine the continuous variables $q_m^s(n)$ and $\bm{P}$ at each hovering position. Specifically, the fitness of each whale agent in the population is assessed (step \ref{l6}). The NDS procedure is employed to categorize the population into distinct PFs (step \ref{l7}). Additionally, the crowding distances are calculated to evaluate the solution diversity (step \ref{l8}).  The best solution $\mathbf{X}^*$, which acts as the leader, is selected based on a combination of non-dominated rank and crowding distance (step \ref{l9}). This selection process ensures that the leader represents a high-quality solution and maintains a diverse set of options for exploration. The main loop iterates over the number of iterations, i.e., $i\in\left[1,I'_{max}\right]$. For each whale agent $j\in\left[1,J_{max}\right]$, the parameters $\vec{A}, \vec{C}, \tau$ and $l$ are iteratively updated (step \ref{l10}).  Then,  the solution $\mathbf{X}^*$ is updated according to different strategies based on the value of $p$ (steps \ref{l11}-\ref{l12}). After updating the positions, the fitness of each whale is re-evaluated. The population is sorted into fronts using the NDS process, and the crowding distances are recalculated. The best solution is selected based on the rank and crowding distance (step \ref{l13}). Repeat the steps  \ref{l6}-\ref{l9} to update solutions. The above process is repeated until the result converges.
Further, the greedy selection mechanism is applied to select the next hovering position $q_m^s(n+1)$ (step \ref{l3}). Then, we alternatively execute NS-WOA and greedy selection mechanism until obtaining the visiting sequence $\phi_m^s$ for T-UAV $v_m^s$.
Finally we obtain the set of NDS $\left\{\bm{\Phi},\bm{P}\right\}$, i.e., the Pareto optimal solutions.

\subsection{Time Complexity Analysis}
The time complexity of Algorithm \ref{A1} is related to the number of GUs $U$ and T-UAVs $M$. The time complexity of K-means method is $\mathcal{O}(I_{max}UM)$ \cite{c20}. The time complexity of constructing the Voronoi diagram with cluster centers is $\mathcal{O}(M^2)$. The time complexity of the Fermat points method is $\mathcal{O}(MN_m^sH^{n}logH^{n})$.
In addition, the time complexity of Algorithm \ref{A2}  is related with three parts: the NS procedure and the WOA algorithm, and the selection greedy mechanism. 
The time complexity of NS is influenced by the population size $J_{max}$ and the number of objective function dimensions $K$, with the worst complexity $\mathcal{O}(K J_{max}^2)$.
Besides, the time complexity of WOA is mainly dependent on the population size $J_{max}$ and the maximum iteration count $I'_{max}$, which is calculated as $\mathcal{O}(N_{m}^{s}J_{max}I'_{max})$. The time complexity of greedy selection mechanism is related to the number of hovering positions $N_m^s$, i.e., $\mathcal{O}(N_{m}^{s}logN_m^s)$. Consequently, we can obtain the time complexity of the INS-WOA as $\mathcal{O}(MN_{m}^{s}logN_{m}^{s}(KJ_{max}^2+N_{m}^{s}J_{max}I'_{max}))$.
 Moreover, as a hybrid algorithm combining WOA with NDS, INS-WOA maintains the proven convergence properties of WOA \cite{c12} while ensuring the Pareto front diversity.

\begin{table}[!t]
  \renewcommand\arraystretch{1.35}
  \begin{center}
     \caption{PARAMETER SETTING}
      {\label{table1}}
      \fontsize{7.8}{10}\selectfont{
      \begin{tabular}{|c|c||c|c|}
          \hline
          Parameter & Value & Parameter & Value \\ 
          \hline 
          $\alpha$ & 9.6& $\beta$ & 0.28\\ 
          \hline
          $f$ & 2.4 GHz & $c$ & $3 \times 10^{8}$ m/s\\
          \hline
          $\mathbf{g}$ & 9.8 m/$\text{s}^2$ & $Q_{u,m}$ &  10 Mb\\
          \hline 
          $P_{u}^{min}$, $P_{m}^{min}$ & 0.001 W & $P_{m}^{max}$ & 5 W \\
          \hline
          $T_u^{max}$ &  0.4 s & $P_{u}^{max}$ & 1 W \\
          \hline 
          $B_{u,m}$ & 1.8 MHz & $B_{m,s}$ & 5 MHz \\
          \hline
          $\mathbb{U}_{tips}$ & 120 m/s  & $W$ & 4.25 kg \\
          \hline
          $A_{0}$ & 0.5 $\text{m}^2$ & $ v_{0}$ & 0.002 m/s \\
          \hline
          $P_{0}$ & 99.66 W  & $P_{1}$ & 120.16 W \\
          \hline
          $\rho_{0}$ &  1.225 $\text{kg/m}^3$ & $d_0$ & 0.48\\
          \hline
          $||\vartheta_{x,y}||$ & 15 $\text{m/s}$ & $||\vartheta_{z}||$ & 6 $\text{m/s}$\\
          \hline
          $s_{0}$ &  0.0001  &  $\sigma ^2 $ & -174 dBm/Hz\\
          \hline
          $\eta_{LoS}$ &  0.1  &  $\eta_{NLoS}$ & 20 \\
          \hline
      \end{tabular}
      }
  \end{center}
\end{table}

\section{Simulation Results and Analyses}\label{Section5}
We conduct extensive simulations in this section to investigate the performance of the proposed algorithms,  using MATLAB R2020b as the simulation platform for validation. The size of the 3D data collection area is set to $2,000$ $\times$ $2,000$ $\times$ 120 $ \text{m}^3$. Moreover, there are a total of 3 UAV swarms, consisting of 3 H-UAVs and 8 T-UAVs. 
The flight spatial range of H-UAVs and T-UAVs are set as $\Theta_s = \left\{\left[0,2,000 \right],\left[0,2,000 \right],120 \right\}$m and $\Theta_m = \left\{\left[0,2,000 \right],\left[0,2,000 \right],\left[30,100 \right]\right\}$m, respectively. Then, the maximum number of served GUs for the T-UAV at one hovering location is limited to  $U_{max}=6$, and each swarm can consist of up to $M_{max}=3$  T-UAVs. 
Furthermore, the maximum tolerated delay $T_{u}^{max}$ is set as 0.5s according to the requirements of GUs.
The major parameters are summarized in Table \ref{table1}.

\begin{figure}[!t]
  \centering
    \includegraphics[width=0.76\linewidth]{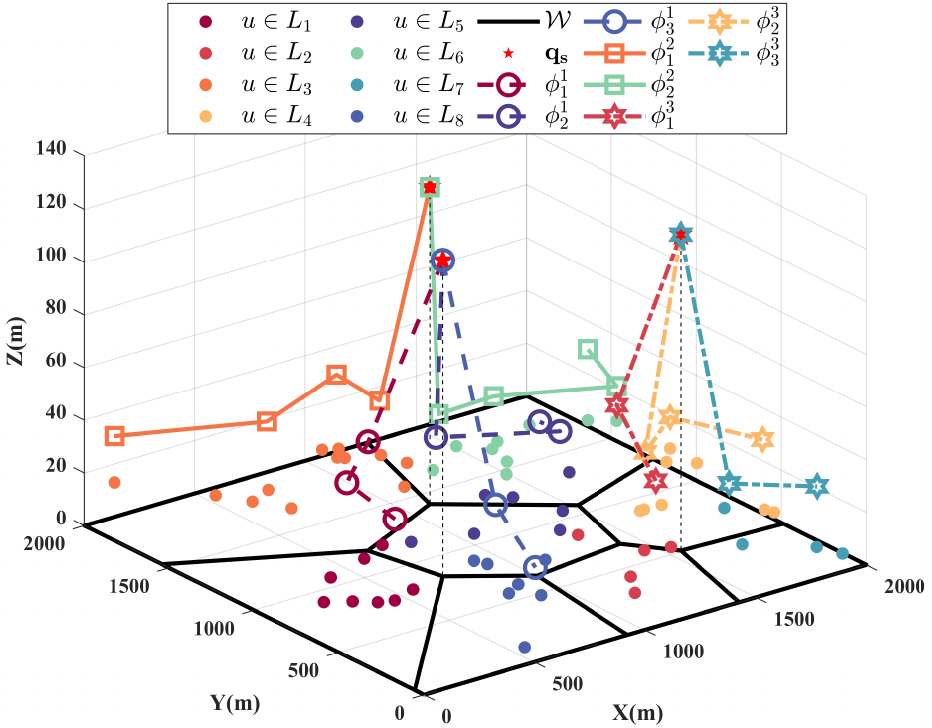}
  \caption{Deployment and trajectory of UAV swarms obtained by INS-WOA with 60 GUs.}
  \label{5}
\end{figure}

The distribution of 60 GUs is depicted in Fig. \ref{3D}. Algorithm \ref{A1} is applied to obtain the deployment positions of UAV swarms, and the number of T-UAVs $M_s$ assigned to the $s$-th UAV swarm as well as the trajectory path are shown in Fig. \ref{5}. It is observed that 3 H-UAVs are deployed at the intersections of Voronoi diagrams, and 8 T-UAVs are dispatched to collect data from GUs.

 We evaluate the proposed INS-WOA  in terms of performances across all three objectives, i.e., TEU, AEG and ADG, by comparing it with other scheduling MOP algorithms, including the  non-dominated sorting genetic algorithm II (NSGA-II) \cite{D2D}, multi-objective grey wolf optimizer (MOGWO) \cite{MOGWO}, and multi-objective artificial hummingbird algorithm (MOAHA) \cite{b33}. 
 In detail, NSGA-II, as a classic multi-objective evolutionary algorithm, has been regarded as a benchmark in the field of multi-objective optimization. 
Derived from the grey wolf optimizer, MOGWO is a powerful multi-objective swarm intelligence optimization algorithm. 
As a novel multi-objective optimization algorithm, MOAHA has attracted attentions in recent years due to its unique foraging behavior simulation and excellent balance between the exploration and exploitation. 
Moreover, with the same size of the UAV swarms, we alter the number $U$ and distribution of GUs to obtain multiple results.

%\begin{figure}[!t]
  %\includegraphics[width=3in]{fig5}
 % \centering
  %\subfloat[ 3D deployment and trajectory of UAV swarms.]{\label{5a}
   %   \includegraphics[width=0.78\linewidth]{O4.pdf}}
  %\quad\quad
  %\subfloat[2D deployment and trajectory of UAV swarms.]{\label{5b}
   % \includegraphics[width=0.73\linewidth]{O5.pdf}}
  %\caption{Deployment and trajectory of UAV swarms obtained by INS-WOA with 60 GUs.}
  %\label{5}
%\end{figure}

\begin{figure}[!t]
  \centering
    \includegraphics[width=0.73\linewidth]{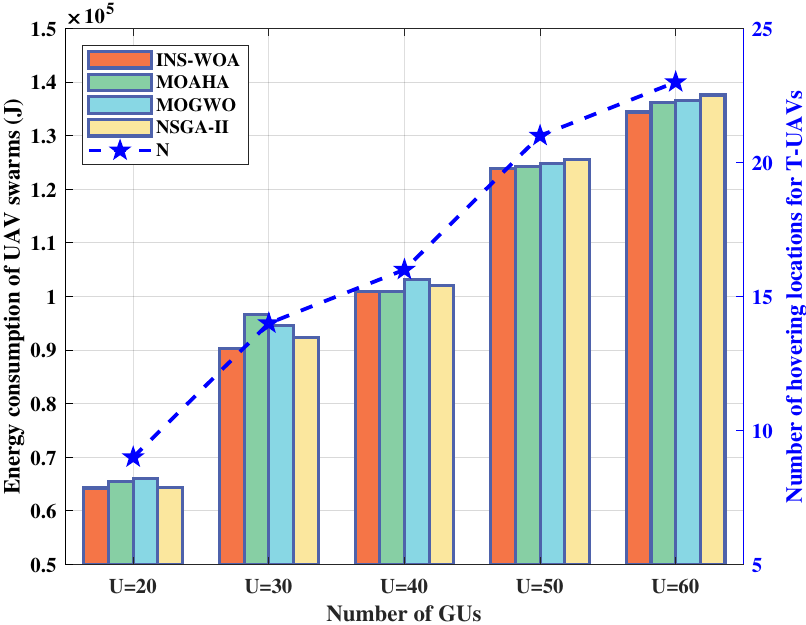}
  \caption{Energy consumption of UAV swarms with different number of GUs.}
  \label{m1}
\end{figure}

Fig. \ref{m1} verifies the total energy consumption of UAV swarms under scenarios with different numbers of GUs for various algorithms. It is obvious that the INS-WOA exhibits a significant performance advantage in terms of TEU solutions. Furthermore, the number of hovering points of T-UAVs $N$ increases with the growth of the number of GUs, which subsequently leads to an increment in the total energy consumption of UAV swarms.

\begin{figure}[!t]
  \vspace{-0.08em}
  \centering
    \includegraphics[width=0.73\linewidth]{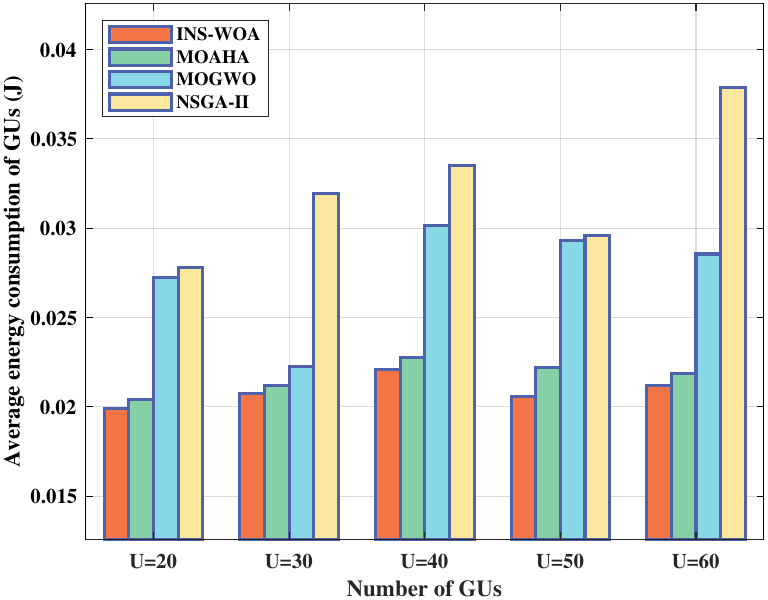}
  \caption{Average energy consumption of GUs with different number of GUs.}
  \label{m2}
\end{figure}
\begin{figure}[!t]
  \vspace{-0.0em}
  \centering
    \includegraphics[width=0.73\linewidth]{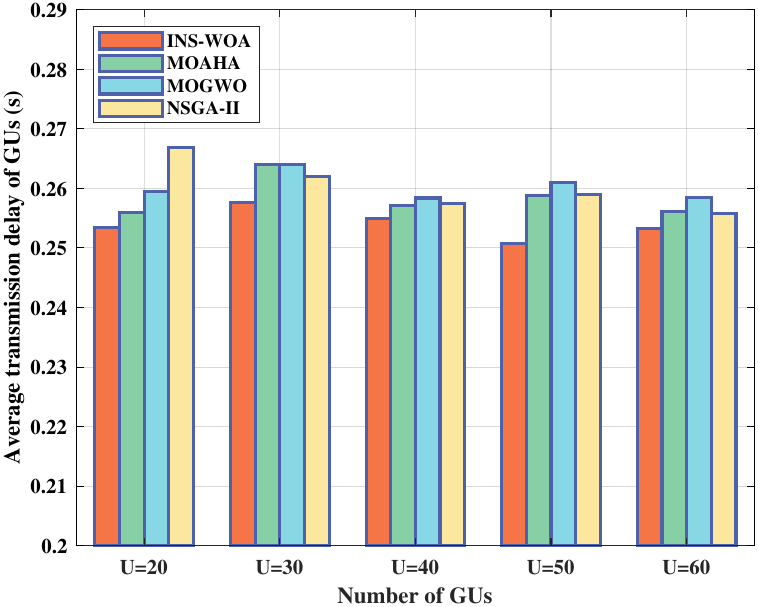}
  \caption{Average transmission delay with different number of GUs.}
  \label{m3}
\end{figure}
\begin{figure}[!t]
  \vspace{-0.01em}
  \centering
    \includegraphics[width=0.73\linewidth]{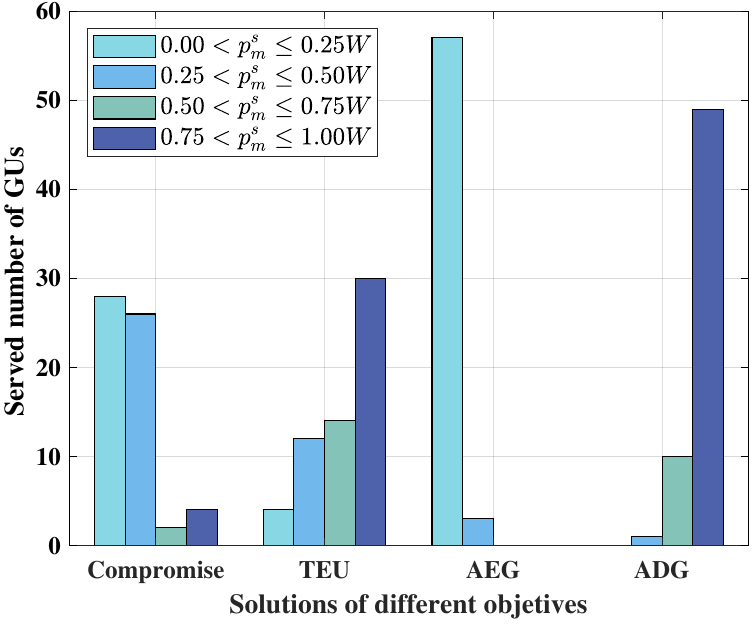}
  \caption{Power control of GUs with different objective solutions. }
  \label{p1}
\end{figure}
\begin{figure}[!t]
  \centering
    \includegraphics[width=0.76\linewidth]{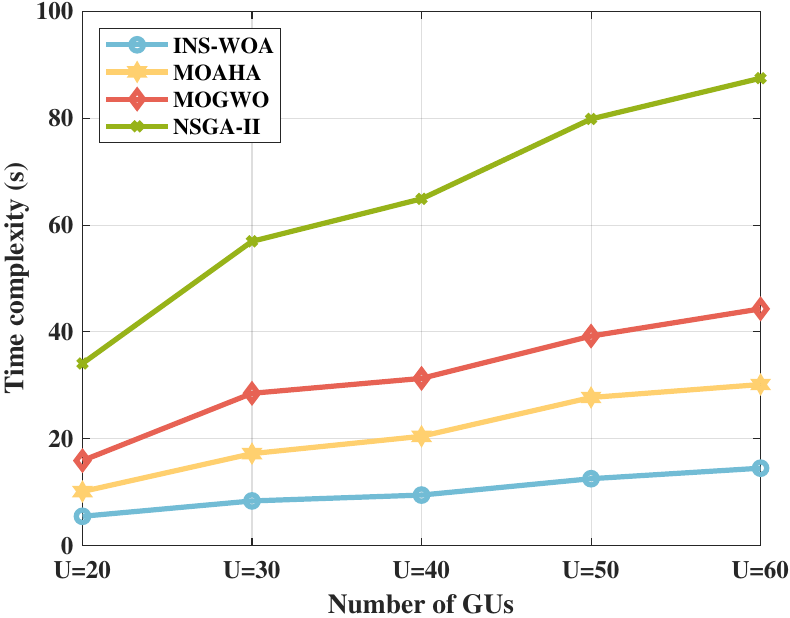}
  \caption{Performance of time complexity under different GUs scales. }
  \label{T1}
\end{figure}

Fig. \ref{m2}  evaluates the average energy consumption of GUs versus different numbers of GUs for various  algorithms. As the number of GUs increases, the energy consumption of GUs in the proposed INS-WOA remains stable at around 0.02 J, which is reduced by 30\% compared with the MOGWO and NSGA-II methods.
This is explained by the fact that INS-WOA  demonstrates a distinct advantage in power control to ensure relatively low energy consumption of GUs.

The average transmission delay of GUs is shown in Fig. \ref{m3}. It is observed that the INS-WOA outperforms other methods in terms of average transmission delay performance under different number of GUs.
Therefore, we obtain the conclusion that the INS-WOA method
has applicability in optimizing average transmission delay and
energy consumption of GUs.

%\begin{figure}[!t]
%  \centering
%    \includegraphics[width=0.76\linewidth]{conv.pdf}
%  \caption{\textcolor{blue}{Comparison of Pareto optimal solutions.} }
%  \label{conv}
%\end{figure}

\begin{figure}[!t]
  \centering
  \subfloat[Energy consumption of UAV swarms.]{\label{com1}
      \includegraphics[width=0.75\linewidth]{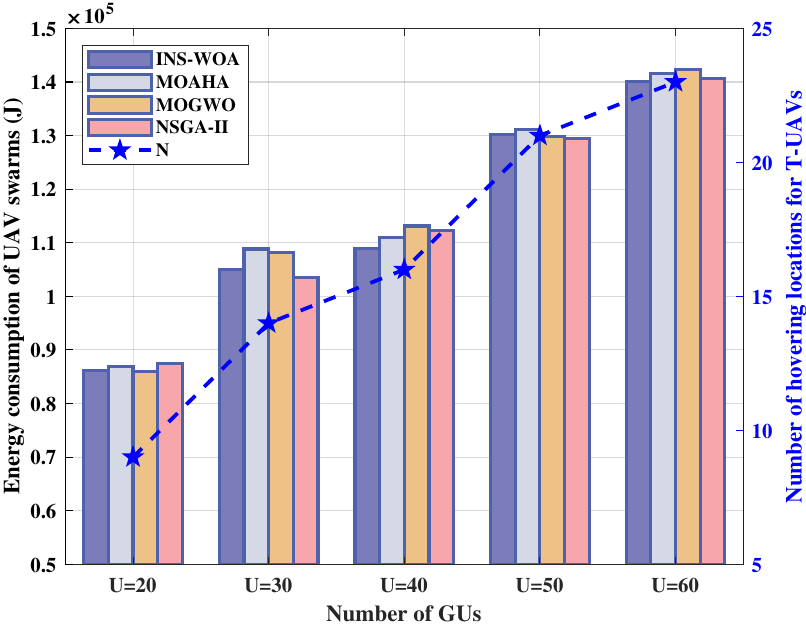}}
  \quad
  \subfloat[Average energy consumption of GUs.]{\label{com2}
    \includegraphics[width=0.72\linewidth]{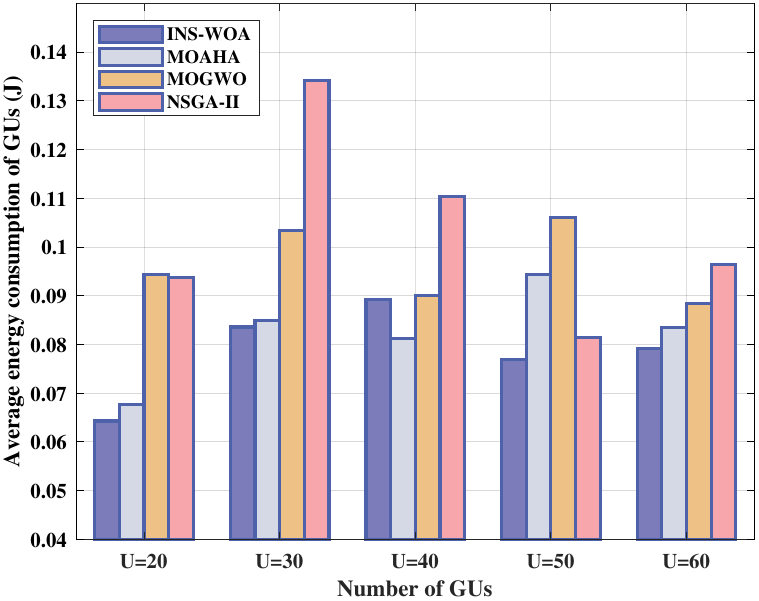}}
  \quad
  \subfloat[Average transmission delay of GUs.]{\label{com3}
    \includegraphics[width=0.72\linewidth]{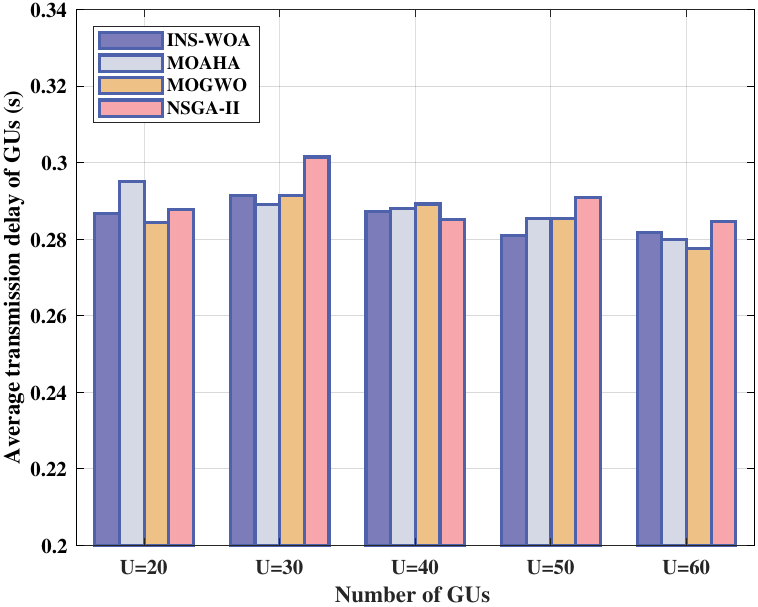}}
  \caption{  Comparison of compromise solutions under different algorithms with 60 GUs.}
  \label{com}
\end{figure}

\begin{figure*}[!t]
  \centering
  \subfloat[ 3D trajectory with the TEU solution.]{
      \includegraphics[width=0.315\linewidth]{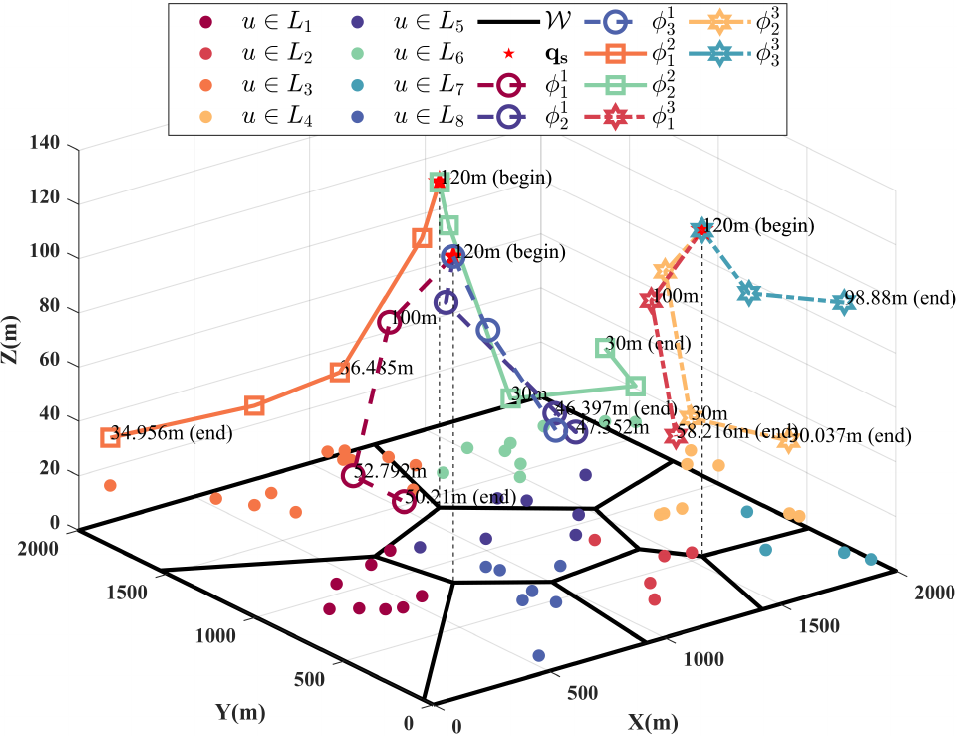}}
  \subfloat[  3D trajectory with the AEG solution.]{
      \includegraphics[width=0.315\linewidth]{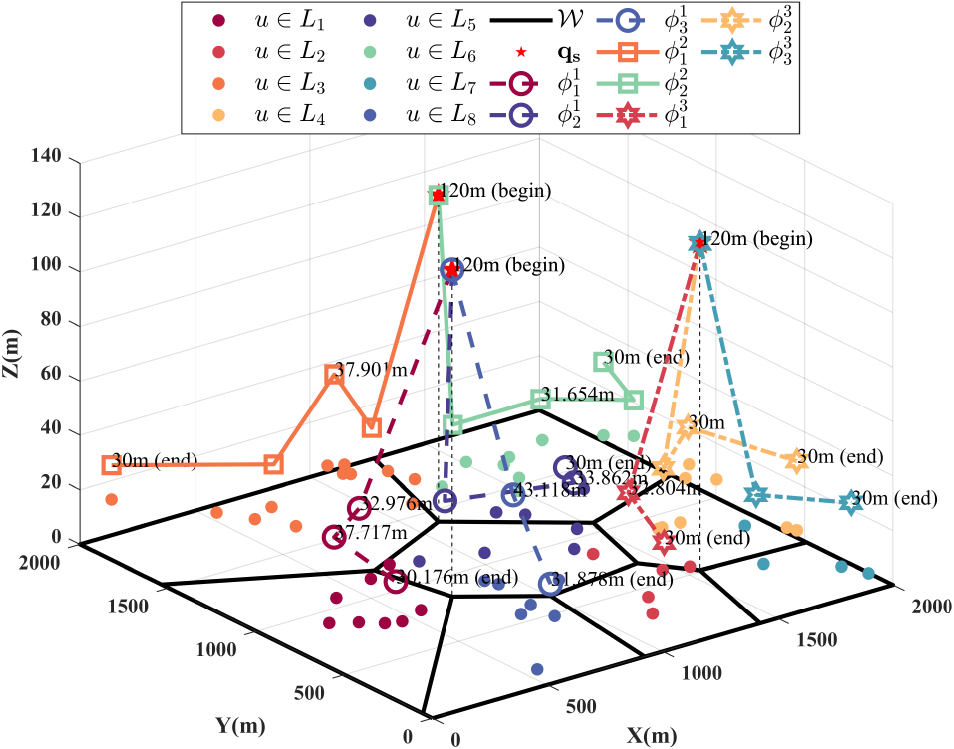}}
  \subfloat[  3D trajectory with the ADG solution.]{
      \includegraphics[width=0.315\linewidth]{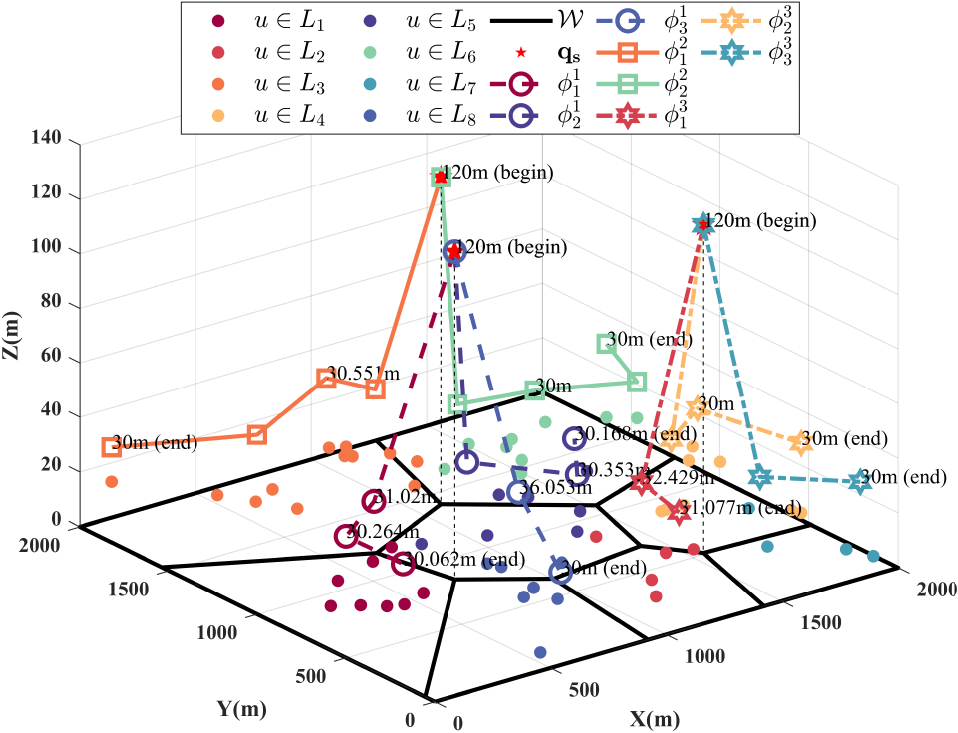}}

  \subfloat[ 2D trajectory with the TEU solution.]{
      \includegraphics[width=0.305\linewidth]{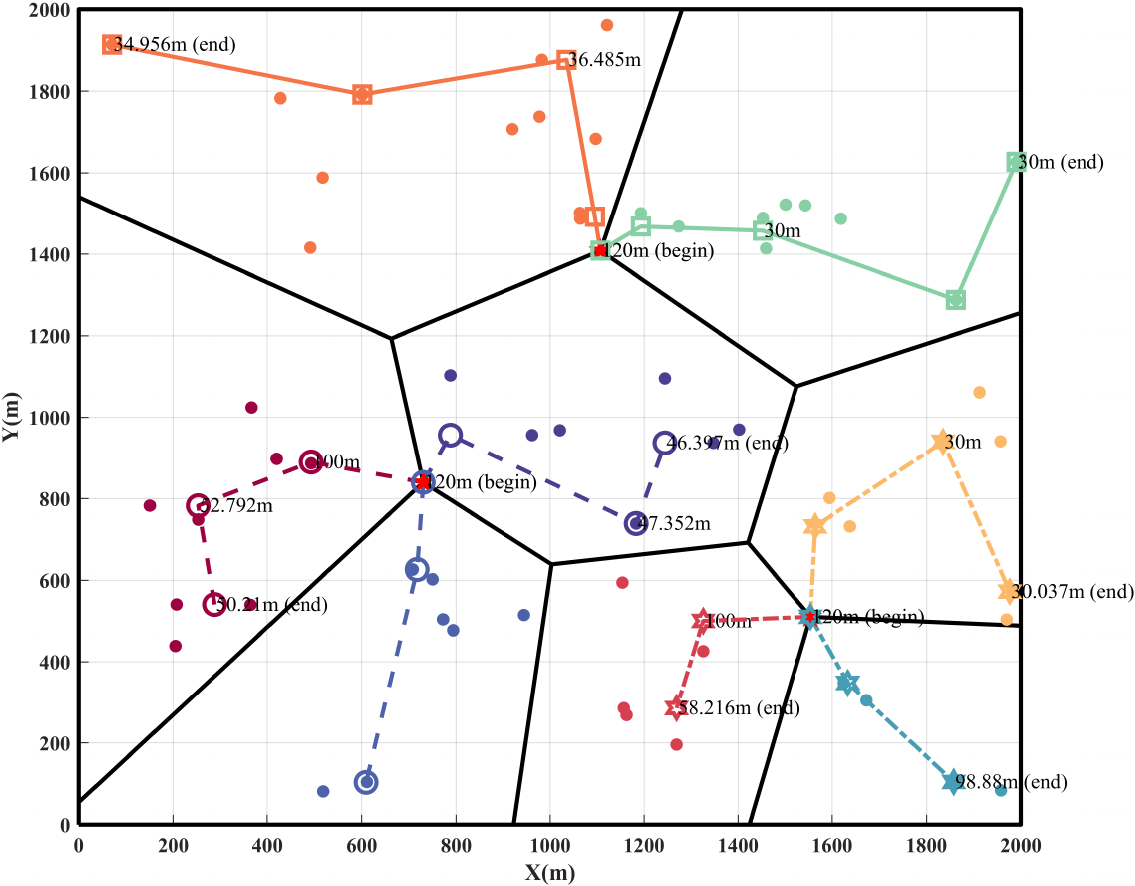}}
  \subfloat[ 2D trajectory with the AEG solution.]{
      \includegraphics[width=0.30\linewidth]{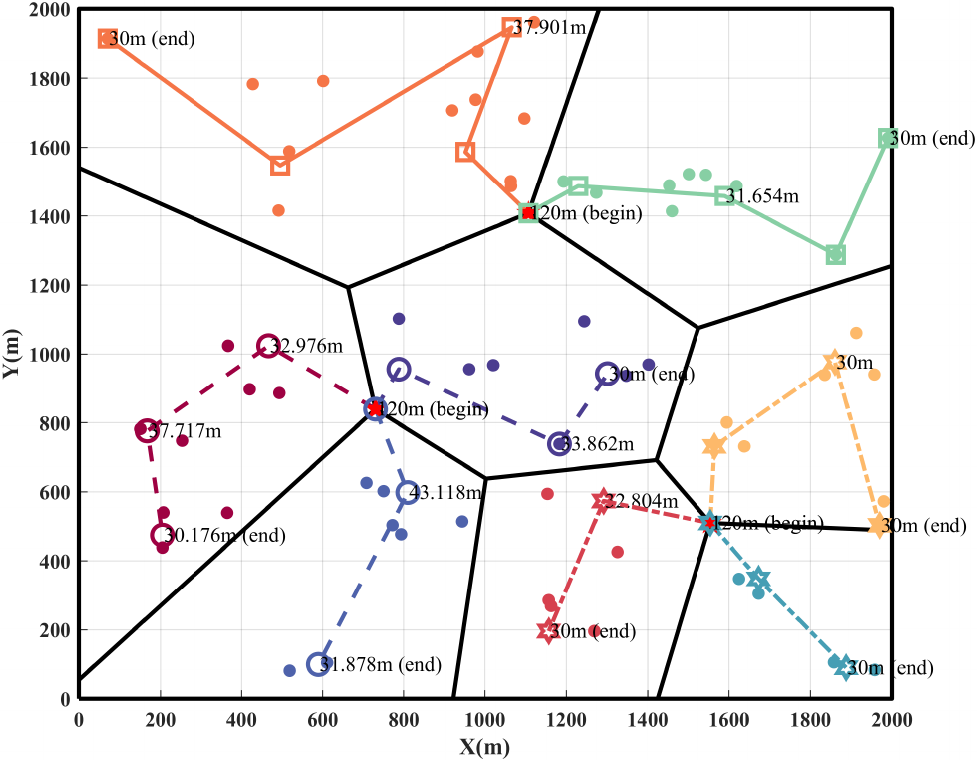}}
  \subfloat[ 2D trajectory with the ADG solution.]{
      \includegraphics[width=0.31\linewidth]{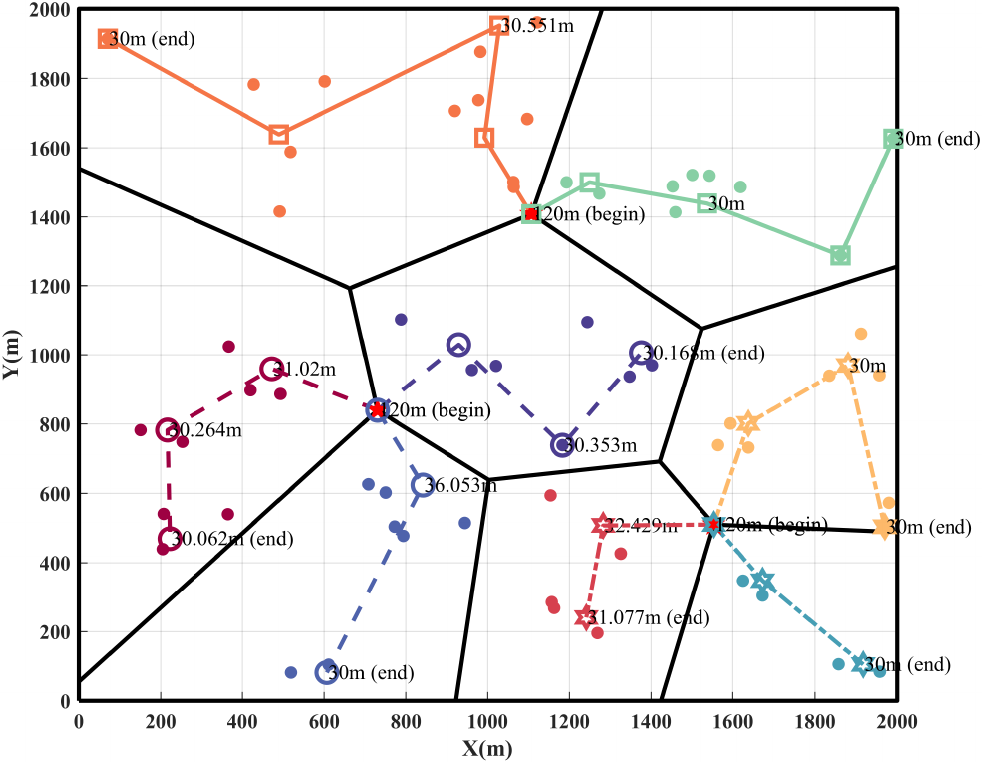}}
  \caption{ Deployment and trajectory of UAV swarms.}
  \label{TEU}
\end{figure*}

Fig. \ref{p1} illustrates the transmission power distribution of 60 GUs under different objective solutions. For the sake of comparison, we standardize and normalize the Pareto solution set, selecting a compromise solution among the three objectives for data analysis. 
Three objectives are respectively normalized to $\left[0,1\right]$ via the min-max scaling method of $B_{norm}=\frac{B-B_{min}}{B_{max}-B_{min}}$, where $B_{norm}$ is the normalized solution, $B$ is the original solution. $B_{max}$ and $B_{min}$ are the maximum and minimal solutions, respectively.
For the compromise solution, the transmission power of the 80\% GUs is controlled within a relatively low range of $\left[0\text{W},0.5 \text{W} \right]$.  Furthermore, for the TEU solution, the distribution of transmission power is relatively dispersed. However, the power distributions of the AEG and ADG solutions show completely opposite trends. 
For the AEG solution considering the energy of GUs, the transmission of 70\% GUs is below $0.25W$. For comparison, the transmission power of 60\% GUs is over $0.75W$ for the ADG solution.
It is explained that  the energy consumption of GUs increases as the transmission power grows, while the transmission delay decreases due to the increasing transmission speed caused by the power increment.

Fig. \ref{T1} provides the time complexity of different algorithms. The INS-WOA algorithm shows obvious superiority by achieving lower time complexity than NSGA-II, MOGWO, and MOAHA.
Moreover, as the scale of GUs increases, the time complexity of INS-WOA grows almost linearly, which provides the possibility for applications in large-scale scenarios. 

%\textcolor{blue}{Furthermore, Fig. \ref{conv} depicts the comparison of Pareto optimal solutions of different methods with the maximum number of iterations. It is evident that the non-dominant solution sets of the four methods approximately form a surface in the 3D solution space,  i.e., Pareto optimal solution frontier, thereby validating the convergence properties.
%}

Considering the trade-off relationship among multiple objectives in MOPs, Fig. \ref{com} presents the results of the three optimization objectives under the compromise solutions of different algorithms. It is observed that the proposed algorithm can achieve relatively satisfied results in all three optimization objectives simultaneously under different scenarios, essentially in the first two objectives illustrated in Fig. \ref{com}(a) and Fig. \ref{com}(b). However, as shown in Fig. \ref{com}(c), for the optimization objective of transmission delay, the INS-WOA method fails to reach the optimal solution. This is accounted for the fact that compared with other algorithms, the compromise solution of the INS-WOA algorithm tends to reduce the energy consumption of UAV swarms and GUs at the cost of transmission delay, so as to achieve the balance and trade-off among multiple objectives. Further, by comparing Fig. \ref{m1} with Fig. \ref{com}(a), we find that the TEU solution is significantly superior to the compromise solution in terms of the energy consumption of UAV swarms. 
Similarly, it is also the case for both the energy consumption and  transmission delay of GUs.

Subsequently, Fig. \ref{TEU} illustrates the deployment and trajectory schematics of the UAV swarms under the three optimization solutions selected from Pareto solutions, respectively.
For TEU solution, T-UAVs tend to hover at higher positions to reduce the path length depicted in Fig. \ref{TEU}(a). This occurs because, when executing the TEU solution, T-UAVs hover at higher altitudes to reduce the flight energy consumption. Consequently, GUs should employ increased transmission power to ensure timely data transmission. In contrast, for both AEG and TEU, T-UAVs are expected to fly close to the GUs at lowest hovering altitude $30$m to decrease the energy consumption and transmission delay, shown in Fig. \ref{TEU}(b) and Fig. \ref{TEU}(c), respectively.

\vspace{-0.5em}
\section{Conclusions}\label{Section6}
In this work, we investigated a hierarchical UAV swarms model for large area data collection in 6G AAN. 
We focused on optimizing the deployment and trajectory of UAV swarms, as well as the power control of GUs and T-UAVs. To tackle the proposed USDC-MOP which simultaneously minimized the total energy consumption of the UAV swarms, the energy consumed by GUs, and the transmission delay of GUs, we proposed two algorithms. Specifically, a pre-deployment method by utilizing the Voronoi diagram and Fermat Points
for UAV swarms was presented to deal the original MOP. Then, we proposed the INS-WOA approach with greedy mechanism to tackle the transformed MOP.  Extensive simulations were carried out to thoroughly assess the performance and effectiveness of the proposed algorithms in three objectives. Compared with other benchmark algorithms, the proposed algorithm significantly reduced  the energy consumption of UAV swarms and transmission delay across different scenarios, with less time complexity.
%In this paper, we investigated and solved the problem of UAV swarms assisted data collection in remote areas. With the aim of minimizing the total energy consumption of the UAV swarms, the average energy consumed by the GU, and the average transmission delay of the GU simultaneously, we formulated the multi-objective optimization problem as the USDC-MOP by jointly optimizing the deployment positions of  UAV swarms, the hovering locations and trajectory of T-UAVs, and the transmission power of T-UAVs and GUs. To solve the formulated MOP, we designed an NS-WOA method with Voronoi diagram area partitioning and Fermat point pre-deployment preprocessing. In particular, by utilizing the partitioned Voronoi regions and the clustering results of GUs, we determined the potential deployment locations for UAV swarms and the number of hovering points for T-UAVs. Subsequently, we the proposed NS-WOA method to jointly optimize the swarm hovering points, the transmission power of T-UAVs, as well as the specific locations of hovering points and their access trajectories. Extensive experiments were carried out to thoroughly assess the performance and effectiveness of the proposed algorithm. When compared with other algorithms, our proposed algorithm significantly reduced both the energy consumption  of the UAV swarm and transmission delay across four  scenarios, while also exhibiting the fastest runtime performance.
\vspace{-0.5em}
\bibliography{ref}
\bibliographystyle{IEEEtran}

\end{document}